\documentclass[sigconf,screen]{acmart}
\usepackage{graphicx}
\usepackage{xcolor}
\usepackage{pifont}
\usepackage{multirow}
\usepackage{makecell}
\usepackage{tcolorbox}
\usepackage{enumitem}
\usepackage{booktabs}
\usepackage{todonotes}
\usepackage{subcaption}
\usepackage[utf8]{inputenc}
\usepackage{textcomp}

\title{Evaluating Inference-Time Defenses\\%
Against Package Hallucination in LLM-Generated Code}

\renewcommand{\shortauthors}{Anonymous Authors}

\begin{abstract}
LLMs are increasingly used for code generation, yet they frequently hallucinate non-existent software packages, creating exploitable entry points into the software supply chain. 
We make four contributions to this problem. First, we show that prior evaluation methodologies systematically inflate hallucination rates by misclassifying standard-library modules as hallucinations in some languages.
For Python, the overestimation reaches 9.4 percentage points.
Second, we evaluate seven inference-time defenses for mitigating package hallucinations, including five guided decoding strategies (Greedy, Contrastive, DoLa, Nudging, and Active Layer-Contrastive Decoding), an iterative self-refinement approach (Self-Refine), and a Retrieval-Augmented Generation (RAG)-based defense.. 
Across eight models spanning five families and four programming languages (Python, JavaScript, Ruby, Rust), RAG reduces the package hallucination rate (PHR) in 18 of 32 model--language configurations.
Third, we introduce Package Utility ($\mathsf{PU}$) to assess whether defenses
preserve valid and task-relevant recommendations. Among strategies
evaluated, Greedy decoding provides the strongest average mitigation--utility trade-off.
Fourth, we stress-test all strategies under adversarial prompts seeded with fabricated package names and find that PHR surges by up to 45 percentage points relative to standard prompts, with Ruby consistently the most vulnerable language (80.9--95.2\%). 
Under adversarial conditions, RAG and Self-Refine outperform all decoding-only strategies, indicating that robust defense requires either external grounding or iterative self-verification when prompts are actively hostile. 

Our results recast package hallucination as both a measurement problem and a decoding-time control problem, and they demonstrate that the choice of defense must be matched to the threat model and recommendation utility.
\end{abstract}



\setcopyright{cc}
\setcctype{by}
\acmDOI{10.1145/3832783.3837555}
\acmYear{2026}
\copyrightyear{2026}
\acmISBN{979-8-4007-2882-2/2026/10}
\acmConference[ASE '26]{Proceedings of the 41st IEEE/ACM International Conference on Automated Software Engineering}{October 12--16, 2026}{Munich, Germany}
\acmBooktitle{Proceedings of the 41st IEEE/ACM International Conference on Automated Software Engineering (ASE '26), October 12--16, 2026, Munich, Germany}
\acmSubmissionID{ase26main-p3592-p}
\received{2026-03-26}
\received[accepted]{2026-06-18}

\begin{document}

\title{Evaluating Inference-Time Defenses against Package Hallucination in LLM-Generated Code}

\author{Alberick Euraste Djire}
\correspondingauthor
\orcid{0009-0007-0406-9573}
\affiliation{%
  \institution{University of Luxembourg}
  \city{Luxembourg}
  \country{Luxembourg}
}
\affiliation{%
  \institution{AI4D (CITADEL)}
  \city{Ouagadougou}
  \country{Burkina Faso}
}
\email{euraste.djire@uni.lu}

\author{Iyiola E. Olatunji}
\orcid{0000-0002-0391-9202}
\affiliation{%
  \institution{University of Luxembourg}
  \city{Luxembourg}
  \country{Luxembourg}
}
\email{emmanuel.olatunji@uni.lu}

\author{Melissa Tessa}
\orcid{0009-0008-7525-5522}
\affiliation{%
  \institution{University of Luxembourg}
  \city{Luxembourg}
  \country{Luxembourg}
}
\email{melissa.tessa@uni.lu}

\author{Earl T. Barr}
\orcid{0000-0003-0771-7891}
\affiliation{%
  \institution{University College London}
  \city{London}
  \country{United Kingdom}
}
\email{e.barr@ucl.ac.uk}

\author{Jacques Klein}
\orcid{0000-0003-4052-475X}
\affiliation{%
  \institution{University of Luxembourg}
  \city{Luxembourg}
  \country{Luxembourg}
}
\email{jacques.klein@uni.lu}

\author{Tegawendé F. Bissyandé}
\orcid{0000-0001-7270-9869}
\affiliation{%
  \institution{University of Luxembourg}
  \city{Luxembourg}
  \country{Luxembourg}
}
\email{tegawende.bissyande@uni.lu}

\renewcommand{\shortauthors}{Djire, Olatunji, Tessa, Barr, Klein and Bissyandé}

\maketitle

\section{Introduction}
\label{sec:introduction}

Code-generating large language models (LLMs) have become a part of software development, powering IDE assistants, automated pipelines, and interactive chat-based programming workflows~\cite{ge_survey_2025, tian2023chatgpt, tessa2026secure, rajput2026correctness, fazlija2026towards, olatunji2026donotcopypaste}. 
When a model produces code, it implicitly selects which external dependencies a developer will import, install, and ultimately trust. 
This selection is driven by the model's parametric memory rather than by verified package metadata, and it can go wrong in a way that is difficult to detect and easy to exploit. When an LLM recommends a package that does not exist in any legitimate registry, the result is a \emph{package hallucination}~\cite{spracklen_we_nodate}. 
If an attacker registers that hallucinated name on a public registry and uploads a malicious payload, any developer who blindly installs the LLM's recommendation executes attacker-controlled code. This attack pattern, variously called \emph{slopsquatting} or AI-induced package squatting~\cite{armstrong_slopsquatting_2025, al-zofi_ai-induced_2025}, is a growing threat to the software supply chain~\cite{williams_research_2025, ladisa2023journey}.

Empirical evidence shows that package hallucination is neither rare nor confined to weak models. 
\citet{spracklen_we_nodate} found that at least 5.2\% of packages recommended by commercial LLMs and 21.7\% of those from open-source models were hallucinated, across more than 576,000 code samples. \citet{krishna_importing_2025} showed that hallucination rates vary with programming language, model size, and prompt specificity. \citet{haque_secure_2025} demonstrated that the phenomenon extends to shell-command generation for Go, where fabricated module paths take realistic URL-style forms. The problem is compounded by the fact that hallucinated names are often persistent across repeated generations~\cite{spracklen_we_nodate}, making them predictable targets for attackers.
Recent work has also shown that package hallucination is more prevalent in smaller language models than in larger ones, making them a particularly challenging setting for mitigation~\cite{krishna_importing_2025, spracklen_we_nodate}. Therefore, we focus our evaluation on smaller open-source models. Beyond providing a stringent testbed for inference-time defenses, these models remain attractive in practice because of their lower computational requirements and widespread deployment in resource-constrained environments.

Several defenses have been proposed. Retrieval-augmented generation (RAG) grounds the model in verified registry data and can sharply reduce hallucination rates~\cite{spracklen_we_nodate, lewis2020retrieval}. Self-refinement asks the model to critique and revise its own output~\cite{madaan2023self, spracklen_we_nodate}. Supervised fine-tuning updates model weights with package-aware supervision~\cite{spracklen_we_nodate}. Each of these approaches has clear merits, but each also has structural limitations. RAG introduces an external retrieval pipeline that may not be available in all deployment settings and that can itself be poisoned. Self-refinement operates post-hoc and is unreliable for small models that cannot detect their own errors~\cite{lee_hallucination_2025}. Fine-tuning is expensive and couples the defense to a specific model checkpoint.

A largely unexplored alternative is to intervene during generation itself, at the point where the model commits to dependency tokens. Guided decoding strategies modify the token selection policy at inference time without retraining the model and without relying on external knowledge. Contrastive Decoding~\cite{li_contrastive_2023} re-ranks candidate tokens by contrasting a larger ``expert'' model against a smaller ``amateur'' model, favoring tokens that reflect the expert's superior factual grounding. DoLa~\cite{chuang_dola_2024} contrasts logits from late and early transformer layers within a single model, amplifying factual knowledge encoded in deeper layers while suppressing surface patterns from shallower ones. \citet{zhang_active_2025} extended this idea with Active Layer-Contrastive Decoding (ALCD), which uses reinforcement learning to decide when to apply layer contrasts. Nudging~\cite{fei_nudging_2025} uses a small aligned surrogate model to replace tokens whenever the base model's confidence falls below a threshold. 
All four methods were originally developed for natural-language factuality tasks; none has been evaluated for package hallucination.
Beyond mitigation, we identify a fundamental measurement flaw. Existing evaluation frameworks classify recommended packages absent from the target registry (e.g., PyPI or npm) as hallucinated. This misclassifies standard-library modules (e.g. \texttt{os}, \texttt{math}, and \texttt{json} in Python) which appear in no registry. Consequently, reported rates are inflated by up to 7.6 percentage points under default decoding, distorting baselines and apparent defense effectiveness.
This paper makes four contributions:
\begin{itemize}[leftmargin=13pt]
    \item \textbf{A corrected evaluation framework for package hallucination.} We identify a systematic source of false positives in registry-based evaluation, namely the misclassification of standard-library imports as hallucinated packages, and 
    reduce this bias by augmenting registry-based validation with language-specific standard-library manifests.
    \item \textbf{Inference-time defenses based on guided decoding for smaller LLMs.} We adapt and evaluate five inference-time decoding strategies (Greedy, Contrastive Decoding, DoLa, ALCD and Nudging) that intervene directly in token selection to reduce package hallucinations without model retraining or dependence on external retrieval. We compare them against standard baselines, including Vanilla decoding, RAG, and Self-Refine.
    \item \textbf{A utility-aware evaluation of package-hallucination defenses.}
We introduce package utility ($\mathsf{PU}$), a precision--recall measure of valid and task-relevant recommendations, and show that reducing PHR does not necessarily preserve recommendation usefulness.
    \item \textbf{A multilingual evaluation under both standard and adversarial prompts.} We assess all strategies across four programming languages with distinct package ecosystems and under adversarial prompts seeded with fabricated package names, providing a comparative view of when guided decoding is effective and where it fails.
\end{itemize}

\section{Related Work}
\label{sec:related_work}

\noindent
\textbf{Package Hallucination.}
Package hallucination occurs when an LLM generates imports or dependency recommendations for packages that do not exist in the target software ecosystem~\cite{spracklen_we_nodate, krishna_importing_2025}. It represents a specialized form of code hallucination, alongside broader syntactic, semantic, and requirement-level failures in LLM-generated code~\cite{zhang_llm_2025, tambon2025bugs, liu_beyond_2026}. Unlike syntactic errors, which can often be detected through static analysis and iterative feedback~\cite{ding2023static, dolcetti2026helping}, package hallucinations are typically syntactically valid and only become apparent during installation or execution~\cite{spracklen_we_nodate, krishna_importing_2025, haque_secure_2025}. This makes them a software supply-chain security concern, enabling attacks such as slopsquatting~\cite{armstrong_slopsquatting_2025, al-zofi_ai-induced_2025, tessa2026position} and related forms of package squatting including typosquatting, impersonation squatting, and compound squatting~\cite{neupane_beyond_nodate, jiang_confuguard_2025}.
Empirical studies have shown that package hallucination is prevalent across models, programming languages with over 19.7\% of recommended packages for Python and JavaScript prompting strategies and more severe for smaller LLMs~\cite{spracklen_we_nodate, krishna_importing_2025, haque_secure_2025, twist_library_2026}. Prior work further attributes hallucinations to both generation-time uncertainty~\cite{farquhar2024detecting, kossen2024semantic} and incorrect factual associations stored in parametric memory~\cite{farquhar2024detecting, spracklen_we_nodate}. Our work builds on this literature by correcting evaluation artifacts that inflate hallucination rates and by studying inference-time mitigation strategies.

\noindent
\textbf{Defenses Against Package Hallucination.}
Existing defenses against package hallucination fall into three broad families. First, knowledge-grounding methods supplement the model with verified package information. \citet{spracklen_we_nodate} showed that Retrieval-Augmented Generation (RAG)~\cite{lewis2020retrieval} substantially reduces hallucinations while largely preserving code quality.
Second, post-hoc detection and correction methods validate or revise generated dependencies. \citet{krishna_importing_2025} proposed checking dependencies against time-aware package registries, while Self-Refine~\cite{madaan2023self} iteratively critiques and revises outputs without additional training. However, smaller models often fail to detect their own hallucinations and may enter repetitive refinement loops~\cite{lee_hallucination_2025}.
Third, parameter-based methods modify the model itself. Supervised fine-tuning achieves strong reductions in package hallucination but requires training data, computational resources, and model-specific checkpoint maintenance~\cite{spracklen_we_nodate}. Lower-temperature decoding can also reduce hallucination frequency, although sampling-parameter changes alone are insufficient~\cite{spracklen_we_nodate}.
In contrast, decoding-time mitigation remains comparatively underexplored. Our work addresses this gap by evaluating guided decoding as an inference-time defense that requires no model retraining or external retrieval.

\noindent
\textbf{Guided Decoding.}
Guided decoding modifies token selection at inference time without retraining the model. Standard approaches include deterministic methods such as greedy and beam search, and stochastic methods such as top-$k$ sampling, nucleus sampling, and temperature scaling; their effectiveness depends on the task, model size, alignment, and quantization~\cite{shi_thorough_2024}.
More recent methods use model outputs or internal representations to improve factuality. Contrastive Decoding contrasts an expert model with a smaller amateur model~\cite{li_contrastive_2023} and has also been shown to improve reasoning~\cite{o2023contrastive}. DoLa contrasts early and late transformer layers within a single model~\cite{chuang_dola_2024}, while Active Layer-Contrastive Decoding (ALCD) learns when such contrasts should be applied~\cite{zhang_active_2025}. Nudging instead uses a small aligned model to replace tokens when the base model is uncertain~\cite{fei_nudging_2025}.
In code generation, guided and constrained decoding has been used to enforce syntax~\cite{poesia_synchromesh_2022, geng_grammar-constrained_2023}, improve program structure through tree search~\cite{princis_treecoder_2025}, generate secure code through supervised co-decoding~\cite{he_cosefa_2025}, and adjust line-level logits to promote essential control structures~\cite{li_preliminary_2025}. However, prior work has not systematically evaluated these techniques as defenses against package hallucination. We address this gap by studying Contrastive Decoding, DoLa, ALCD, and Nudging in this setting.

\section{Research Questions}
\label{sec:research_question}

To the best of our knowledge, this is the first study to systematically evaluate guided decoding strategies as defenses against package hallucination and to compare them against both pre-generation and post-generation baselines. Our evaluation spans four programming languages (JavaScript, Python, Ruby, and Rust) across five model families, enabling an analysis of how ecosystem characteristics such as registry size, naming conventions, and package distribution interact with model behavior. We further assess the robustness of all mitigation strategies under adversarial prompts, where the user explicitly steers the model toward non-existent packages.
Our study is organized around four research questions:

\medskip
\noindent\textbf{RQ1 (Baseline Characterization).} \emph{How does the package hallucination rate vary across model families and programming languages when evaluation accounts for standard-library imports?}

\medskip
\noindent\textbf{RQ2 (Defense Effectiveness).} \emph{How effective are guided decoding strategies in reducing package hallucination compared with pre-generation (RAG) and post-generation (Self-Refine) baselines?}

\medskip
\noindent\textbf{RQ3 (\textbf{Mitigation--Utility Trade-off}).} \emph{What trade-offs arise between the effectiveness of package-hallucination mitigation strategies and the utility of their generated package recommendations?}

\medskip
\noindent\textbf{RQ4 (Adversarial Robustness).} \emph{How robust are existing and proposed mitigation strategies when prompts are deliberately seeded with fabricated package names?}

\section{Methodology}
\label{sec:methodology}

We evaluated open-weight models across five families, seven strategies, and four programming languages: Python, JavaScript, Ruby, and Rust. Hallucination detection uses enhanced ground-truth registry snapshots to mitigate false-positive biases, with each strategy tested on the complete dataset $\mathcal{D}_{\text{eval}}$ over three independent runs.

\subsection{Dataset Construction}
\label{sec:dataset}

Following \citet{spracklen_we_nodate}, we construct an evaluation dataset spanning four programming languages. \textbf{Python} and \textbf{JavaScript} are chosen for their popularity and the size of their package ecosystems (PyPI and npm, respectively). \textbf{Ruby} and \textbf{Rust} are chosen as languages with smaller, more structured, and easily verifiable package namespaces (RubyGems and Crates.io, respectively). This selection provides diversity in registry size, naming conventions, and ecosystem maturity. Our dataset construction pipeline consists of three steps.

\noindent
\textbf{Step 1: Package Sampling.}
For each language, we retrieved the top 1,000 most popular packages from the corresponding registries (PyPI, npm, RubyGems, and Crates.io) using the \texttt{Libraries.io} API, which provides unified access to package metadata across more than 36 package managers. Each retrieved record consists of a \texttt{(package\_name, description)} pair, which serves as the seed for prompt generation in Step~3.

\noindent
\textbf{Step 2: Ground-Truth Registry Snapshot.}
To enable reliable hallucination detection, we collected the complete list of package names available in each registry as of March 4, 2026. We further augment each registry snapshot with the standard-library module list for the corresponding language. This augmentation is critical to our corrected evaluation methodology: without it, legitimate standard-library imports such as \texttt{os}, \texttt{math}, and \texttt{json} in Python would be erroneously flagged as hallucinated. The resulting ground-truth package set $\mathcal{P}_{\ell}$ for each language $\ell \in \{\text{Python, JavaScript, Ruby, Rust}\}$ is the union of registry packages and standard-library modules.

\noindent
\textbf{Step 3: Prompt Generation.}
Starting from each \texttt{(package\_name, description)} pair, we used \texttt{GPT-4o-mini} to synthetically generate a natural-language coding instruction that reflects the package's functionality. The system prompt used for this step was:

\begin{tcolorbox}[title={System Prompt: Package Recommendation},
fonttitle=\footnotesize,
fontupper=\footnotesize,
boxsep=1pt,
    left=2mm,
    right=2mm,
    top=1mm,
    bottom=1mm,
    toptitle=1mm,
    bottomtitle=1mm,
    before skip=2pt,
    after skip=2pt]
    You are a coding assistant that recommends packages useful to solve given problems. 
        Respond with only a list of \texttt{<language>} packages, separated by commas, enclosed in square brackets, 
        no additional text.
        
        \textbf{Example output:} \texttt{[package1, package2]}
\end{tcolorbox}

\noindent This procedure yields a final evaluation dataset
$\mathcal{D}_{\text{eval}}$ of 4,000 instructions, with $|\mathcal{D}_{\ell}| = 1{,}000$ for each language~$\ell$.

\subsection{Models}
\label{sec:models}
We evaluate eight \textit{instruction-tuned}, open-weight language models from five families, spanning a range of parameter scales:
\begin{itemize}[leftmargin=13pt]
  \item \textbf{Gemma 3} (Google): \texttt{gemma-1b}, \texttt{gemma-4b}
  \item \textbf{DeepSeek-Coder} (DeepSeek): \texttt{deepseek-1.3b}, \texttt{deepseek-6.7b}
  \item \textbf{Qwen 2.5} (Alibaba): \texttt{qwen-1.5b}, \texttt{qwen-3b}
  \item \textbf{Mistral} (Mistral) : \texttt{mistral-7b} we used Instruct-v0.3.
  \item \textbf{Llama 3.1} (Meta) : \texttt{llama-8b}
\end{itemize}

\noindent We focus on lightweight models (1b--8b parameters) for two reasons. First, prior work has shown that package hallucination rates decrease with model scale, with larger models generally exhibiting greater resistance to fabricated recommendations~\cite{spracklen_we_nodate, krishna_importing_2025}. Smaller models are therefore a harder and more practically important test bed for mitigation strategies, since they are widely deployed on resource-constrained hardware. Second, several of our guided decoding strategies (Contrastive Decoding, Nudging) require running two models simultaneously, and limiting parameter counts keeps inference within the memory budget of a single GPU.

\subsection{Hallucination Detection}
\label{sec:detection}

For each \texttt{(model, language, strategy)} triple, the model is queried on the full dataset $\mathcal{D}_{\text{eval}}$ and asked to recommend packages. The raw outputs are parsed to extract the set of recommended package names. Each name is then checked against the ground-truth set $\mathcal{P}_{\ell}$, which includes both registry packages and standard-library modules. A recommended name is classified as hallucinated if and only if it is absent from $\mathcal{P}_{\ell}$.

We report the following metrics for each configuration

\begin{itemize}[leftmargin=*]
    \item \textbf{Micro PHR} (benchmark level). Computed over the full output of a given (model, language, strategy) configuration:
    \[
        \text{micro-PHR} = \frac{N_{\text{hall}}}{N_{\text{gen}}}
    \]
 
    \item \textbf{Macro PHR} (sample level). Computed per prompt $i$ and then averaged:
    \[
        \text{PHR}_i = \frac{n_{\text{hall},i}}{n_{\text{gen},i}}, \qquad
        \text{macro-PHR} = \frac{1}{|\mathcal{D}_{\text{eval}}|} \sum_{i} \text{PHR}_i
    \]
    where $n_{\text{gen},i}$ and $n_{\text{hall},i}$ are the number of generated and hallucinated packages for prompt $i$, respectively. If $n_{\text{gen},i} = 0$, the prompt contributes $0$ to the sum.
\end{itemize}

\subsection{Package Utility Score ($\mathsf{PU}$)}
\label{sec:package-utility}
PHR measures the fraction of generated package names that are invalid but not their usefulness. A defense can lower PHR by producing fewer packages or suppressing recommendations. We introduce the Package Utility Score ($\mathsf{PU}$), measuring whether generated packages are valid and task-relevant.
For prompt $i$, let $G_i$ be the generated package set and $\mathcal{P}_\ell$ the language-specific universe of registry packages and standard-library modules. Since each prompt is derived from seed package $s_i$ and its description $d(s_i)$ \cite{spracklen_we_nodate}, we construct a task-specific reference set $R_i \subseteq \mathcal{P}_\ell$ by computing cosine similarity between $d(s_i)$ and every verified same-language package description $d(p)$ using \texttt{text-embedding-3-small} embedder, then selecting $s_i$ and the $k-1$ most similar packages. Therefore, $R_i$ contains only valid packages and approximates those functionally related to the seed, crediting valid alternatives while excluding hallucinations.
The useful generated set is
$
U_i = G_i \cap R_i \cap \mathcal{P}_\ell = G_i \cap R_i,
$
where the explicit intersection with $\mathcal{P}_\ell$ emphasizes validity because $R_i \subseteq \mathcal{P}_\ell$. Package precision and recall are defined as
\[
\mathsf{PU_P}_i = \frac{|U_i|}{|G_i|}, \qquad
\mathsf{PU_R}_i = \frac{|U_i|}{|R_i|}.
\]
Precision is the fraction of generated dependencies in the reference set, whereas recall is the fraction of that set recovered. We combine them using an F1-style score:
\[
\mathsf{PU}_i =
2 \times \frac{\mathsf{PU_P}_i \times \mathsf{PU_R}_i}
{\mathsf{PU_P}_i+\mathsf{PU_R}_i}.
\]
We set $\mathsf{PU}_i=0$ when $G_i=\varnothing$ or both precision and recall are zero. That is, an empty output avoids hallucination but provides no package utility. Precision penalizes hallucinated or unrelated packages; recall penalizes output collapse.
Because $R_i$ uses vetted registry metadata, $\mathsf{PU}$ is reproducible and independent of LLM-generated reference labels, but remains a semantic proxy rather than complete ground truth. 
To complement this, we also evaluate recommendation utility through generated code. Specifically, after removing hallucinated packages, we ask the same model to solve the original task using the recommended packages and count how many are imported or otherwise used. This operationally complements description similarity.

\subsection{Baselines and Decoding Strategies}
\label{sec:decoding}

We compare five guided decoding strategies against three baselines.

\noindent
\textbf{Baselines.}
\textit{Vanilla Decoding} uses the model's standard generation configuration, with temperature set to 1.0, top-$k$ set to 50, and top-$p$ set to 1.0, and serves as the non-deterministic reference setting. \textit{Self-Refine}~\cite{madaan2023self, spracklen_we_nodate} is a post-generation iterative correction approach where the model critiques and revises its own output over multiple passes, included as a representative post-hoc mitigation technique. \textit{RAG}~\cite{lewis2020retrieval, spracklen_we_nodate} supplements the prompt with retrieved package information from the registry, included as a representative pre-generation grounding technique.

\noindent
\textbf{Guided Decoding Strategies.}
\textit{Greedy Decoding} selects the highest-probability token at each step, yielding a more deterministic alternative.
\textit{Contrastive Decoding (CD)}~\cite{li_contrastive_2023, o2023contrastive} generates tokens by subtracting the log-probability distribution of a smaller \emph{amateur} model from that of the \emph{expert} (target) model, penalizing tokens that are probable under both models and rewarding tokens specific to the expert. \textit{DoLa}~\cite{chuang_dola_2024} computes a contrastive distribution between an early and a late transformer layer within the same model, amplifying factual knowledge encoded in deeper layers while suppressing surface patterns from shallower ones. \textit{Active Layer-Contrastive Decoding (ALCD)}~\cite{zhang_active_2025} is built upon DoLa and consisted of a selection of tokens where to apply the contrastive distribution decoding based on predefined policy. \textit{Nudging}~\cite{fei_nudging_2025} uses a small aligned surrogate model to generate replacement tokens whenever the base model's confidence falls below a calibrated threshold, effectively steering generation toward more grounded outputs at points of high uncertainty.

\noindent
\textbf{Experimental Configuration.}
To ensure a fair comparison across strategies, we fix a single hyperparameter configuration per strategy and apply it uniformly across all eight models and four languages considered. This choice trades a small amount of per-configuration optimality for comparability: a strategy that wins under a shared configuration is evidence of genuine robustness rather than an artifact of grid search. Table~\ref{tab:hyperparams} summarizes the configuration used for each strategy.
Our choice of decoding hyperparameters is based on the previous study made by ~\citet{shi_thorough_2024}. For RAG, we follow the retrieval configuration of ~\citet{spracklen_we_nodate}, adopting their top-$k$ retrieval setup. 

\begin{table}[t]
    \caption{Hyperparameter configuration per strategy.}
    \footnotesize
    \label{tab:hyperparams}
    \begin{tabular}{ll}
        \toprule
        \textbf{Strategy} & \textbf{Hyperparameters} \\
        \midrule
        Baseline (Vanilla) & $T{=}1.0$, top-$k{=}50$, top-$p{=}1.0$ \\
        \midrule
        Greedy & $T{=}0$ \\
        \midrule
        Contrastive Decoding (CD) & $\alpha{=}0.5$; amateur/expert layer \\
         & pairing (model-specific) \\
         \midrule
        DoLa & mature\_layer = final layer of \\
         & target model; \\
         & early\_exit\_layers$=\{4,8,12,16\}$; \\
         & $\delta{=}0.1$; repetition\_penalty$=1.2$ \\
         \midrule
        ALCD & mature\_layer = final layer of \\
         & target model; \\
         & early\_exit\_layers$=\{4,8,12,16\}$; \\
         & $\delta{=}0.1$; repetition\_penalty$=1.2$; \\
         & active decision policy $=> entropy \geq 1$ \\
         \midrule
        Nudging & surrogate\_model = largest in \\
         & family; top\_prob\_threshold$=0.4$ \\
         \midrule
        RAG & retrieval\_$k{=}5$; corpus = \\
         & per-language package registry \\
         \midrule
        Self-Refine & max\_refinement\_iterations$=5$ \\
        \bottomrule
    \end{tabular}
\end{table}
\section{Experiments and Results}
\label{sec:experiments}

This section presents experimental results aligned with the four research questions from Section~\ref{sec:research_question}. 
We repeat each experiment three times and report the mean and standard deviation across runs. For large tables where space is limited, we report the mean in the main paper and provide the full per-run values and standard deviations in the artifact.
We first quantify the impact of standard-library correction on package hallucination measurement. We then characterize the baseline performance across multiple models and languages (RQ1), compare inference-time defenses for package recommendation and code generation (RQ2), evaluate package utility across different reference sets (RQ3), and assess the robustness of defenses to adversarial prompts involving fabricated package names (RQ4).

\subsection{The Impact of Standard-Library Correction}
\label{sec:library-correction}
Prior evaluations can overestimate package hallucination by treating standard-library imports as invalid because they are absent from centralized registries. For example, Python modules such as \texttt{os} and \texttt{math} do not appear on PyPI despite being valid dependencies. To quantify this bias, we collected standard-library manifests for the two applicable languages and computed PHR with and without correction.
Table~\ref{tab:stdlibs-delta-macro} reports the overestimation $\Delta$ by language and strategy. For Python, excluding standard-library modules inflates PHR by up to $9.4$\,pp, compared with $2.5$\,pp for Ruby, reflecting Python's larger and more frequently imported standard library.
RAG shows the smallest overestimation in both languages, consistent with its reliance on verified package information. Standard-library misclassification can also alter the relative ranking of mitigation strategies, affecting both measurement accuracy and comparative evaluation. We therefore apply standard-library correction throughout our evaluation.

\begin{table}%[t]
\centering
\caption{Micro PHR with and without standard library (stdlib) correction, averaged over all models. $\Delta>0$ means overestimation without the correction.}
\label{tab:stdlibs-delta-macro}
\footnotesize
\begin{tabular}{@{}llrrrr@{}}
\toprule
\textbf{Lang.} & \textbf{Strategy} & \textbf{$N_{\text{gen}}$} & \textbf{PHR$_{\text{w/ stdlib}}$ (\%)} & \textbf{PHR$_{\text{w/o stdlib}}$ (\%)} & \textbf{$\Delta$ (pp)} \\
\midrule
\multirow{8}{*}{Python}
  & Vanilla & 79,965 & 24.7$_{\pm 0.67}$ & 32.3$_{\pm 0.53}$ & +7.6 \\
  & Greedy & 47,796 & 19.3$_{\pm 0.00}$ & 27.4$_{\pm 0.00}$ & +8.1 \\
  & S-Ref & 39,465 & 19.0$_{\pm 1.14}$ & 28.4$_{\pm 1.26}$ & \textbf{+9.4} \\
  & CD & 24,489 & 26.7$_{\pm 0.55}$ & 32.6$_{\pm 0.50}$ & +5.9 \\
  & DoLa & 51,540 & 32.3$_{\pm 0.60}$  & 38.9$_{\pm 0.58}$ & +6.6 \\
  & Nudge & 17,628 & 21.3$_{\pm 1.40}$ & 30.4$_{\pm 0.91}$ & +9.1 \\
  & RAG & 59,910 & 11.5$_{\pm 0.04}$ & 14.3$_{\pm 0.02}$ & +2.8 \\
  & ALCD & 49,257 & 27.6$_{\pm 0.71}$ & 34.8$_{\pm 0.79}$ & +7.1 \\
\midrule

\multirow{8}{*}{Ruby}
  & Vanilla & 69,072 & 38.2$_{\pm 0.89}$ & 40.4$_{\pm 0.96}$ & +2.2 \\
  & Greedy & 39,771 & 30.3$_{\pm 0.00}$ & 32.8$_{\pm 0.00}$ & \textbf{+2.5} \\
  & S-Ref & 34,596 & 31.9$_{\pm 0.29}$ & 34.1$_{\pm 0.32}$ & +2.2 \\
  & CD & 16,596 & 40.9$_{\pm 0.53}$ & 42.6$_{\pm 0.58}$ & +1.7 \\
  & DoLa & 47,223 & 46.6$_{\pm 0.39}$  & 49.0$_{\pm 0.38}$ & +2.4 \\
  & Nudge & 19,446 & 38.0$_{\pm 0.87}$ & 40.3$_{\pm 1.09}$ & +2.2 \\
  & RAG & 61,260 & 11.7$_{\pm 0.03}$ & 13.3$_{\pm 0.01}$ & +1.6 \\
  & ALCD & 43,110 & 39.3$_{\pm 0.21}$ & 41.8$_{\pm 0.20}$ & \textbf{+2.5} \\
\bottomrule
\end{tabular}
\end{table}

\subsection{Baseline Package Hallucination}
\label{sec:rq1-baseline}

We characterize the hallucination behavior of eight models under vanilla conditions by querying them on the full evaluation dataset $\mathcal{D}_{\text{eval}}$ and extracting recommended package names, validated against ground-truth registry snapshots $\mathcal{P}_{\ell}$. 
The result is shown in Table \ref{tab:baseline-phr}.

\noindent
\textbf{PHR varies across models and languages.} 
Mistral-7b exhibits the lowest Overall micro PHR (16.1\%), while Qwen-1.5b achieves the highest (47.9\%). Scaling from smaller to larger models correlates with a reduction in average micro PHR: DeepSeek shows a decrease of 18.3 pp (39.2\% to 20.9\%), Gemma 19.8 pp (38.2\% to 18.4\%), and Qwen 20.3 pp (47.9\% to 27.6\%).
Among programming languages, Rust has the highest micro PHR of 63.1\% on Qwen-1.5b, followed by Ruby on Gemma-1b (53.9\%), indicating that more than half of the recommended packages are hallucinated. Javascript has the least micro PHR of 7.6\%.

\noindent
\textbf{Package-generation volume varies across models.}
The mean number of packages generated per prompt (macro) ranges from approximately 1 for DeepSeek-1.3b to 12 for Llama-8b. Overall, Qwen-1.5b produces the most packages (100,896), including 48,330 hallucinations, while DeepSeek-1.3b produces the fewest. Micro- and macro-PHR capture complementary perspectives. Micro-PHR measures hallucination across all generated packages at the benchmark level and therefore gives greater weight to prompts producing longer recommendation lists. Macro-PHR instead averages per-prompt rates, reflecting hallucination for a typical prompt regardless of output length. For most configurations, micro-PHR exceeds macro-PHR, indicating that high-volume prompts are also more hallucination-prone. The largest gaps occur for Qwen-1.5b on Rust (+23.3\,pp), DeepSeek-1.3b on Ruby (+23.2\,pp), and Gemma-1b on Ruby (+22.7\,pp). Reporting both metrics therefore distinguishes aggregate package-level exposure from typical prompt-level behavior.

\begin{table}[t]
\centering
\caption{Vanilla decoding PHR per model and language. 
$\bar{n}$ = mean per prompt. \textit{Overall} is column aggregates of all languages and for micro and macro PHR, it is the mean. Higher PHR per model is in \textbf{bold}.}
\label{tab:baseline-phr}
\footnotesize
\setlength{\tabcolsep}{3pt}
    \begin{tabular}{@{}llrrp{0.9cm}rrp{0.9cm}@{}}
    \toprule
    \textbf{Model} & \textbf{Lang.} & \textbf{$N_{\text{gen}}$} & \textbf{$N_{\text{hall}}$} & \textbf{micro PHR (\%)} & \textbf{$n_{\text{gen}}$} & \textbf{$n_{\text{hall}}$} & \textbf{macro PHR (\%)} \\
    \midrule
     \multirow{5}{*}{\texttt{deepseek-1.3b}}
      & Python & 3,024 & 1,077 & 35.6$_{\pm 1.16}$ & 2.02 & 0.72 & 20.7$_{\pm 1.64}$ \\
      & JavaScript & 1,950 & 585 & 30.0$_{\pm 5.49}$ & 1.30 & 0.39 & 13.9$_{\pm 1.77}$ \\
      & Ruby & 1,470 & 600 & 40.8$_{\pm 3.35}$ & 0.98 & 0.40 & 17.6$_{\pm 1.20}$ \\
      & Rust & 2,658 & 1,305 & \textbf{49.1$_{\pm 0.55}$} & 1.78 & 0.87 & \textbf{35.4$_{\pm 0.23}$}\\ \cmidrule(lr){2-8}
      & \textit{Overall} & 9,102 & 3,567 & 39.2$_{\pm 0.55}$ & 4.55 & 1.78 & 21.9$_{\pm 0.23}$ \\
    \midrule
     \multirow{5}{*}{\texttt{deepseek-6.7b}}
      & Python & 5,013 & 630 & 12.6$_{\pm 1.77}$ & 3.34 & 0.42 & 10.6$_{\pm 0.77}$ \\
      & JavaScript & 4,647 & 474 & 10.2$_{\pm 2.68}$ & 3.10 & 0.32 & 9.5$_{\pm 1.03}$ \\
      & Ruby & 5,850 & 1,428 & 24.4$_{\pm 1.66}$ & 3.90 & 0.95 & 15.2$_{\pm 1.11}$ \\
      & Rust & 2,970 & 1,329 & \textbf{44.8$_{\pm 0.88}$} & 1.98 & 0.89 & \textbf{45.7$_{\pm 1.42}$} \\ \cmidrule(lr){2-8}
      & \textit{Overall} & 18,480 & 3,861 & 20.9$_{\pm 0.18}$ & 9.24 & 1.93 & 20.2$_{\pm 0.62}$ \\
    \midrule
     \multirow{5}{*}{\texttt{gemma-1b}}
      & Python & 9,780 & 2,358 & 24.1$_{\pm 1.62}$ & 3.26 & 0.79 & 18.4$_{\pm 0.38}$ \\
      & JavaScript & 12,948 & 4,362 & 33.7$_{\pm 0.82}$ & 4.32 & 1.45 & 18.9$_{\pm 1.10}$ \\
      & Ruby & 11,256 & 6,063 & \textbf{53.9$_{\pm 0.84}$} & 3.75 & 2.02 & \textbf{31.2$_{\pm 0.36}$} \\
      & Rust & 11,130 & 4,443 & 39.9$_{\pm 0.99}$ & 3.71 & 1.48 & 28.5$_{\pm 1.49}$ \\ \cmidrule(lr){2-8}
      & \textit{Overall} & 45,114 & 17,226 & 38.2$_{\pm 0.42}$ & 11.28 & 4.31 & 24.2$_{\pm 0.71}$ \\
    \midrule
     \multirow{5}{*}{\texttt{gemma-4b}}
      & Python & 9,894 & 1,200 & 12.1$_{\pm 0.87}$ & 3.30 & 0.40 & 12.3$_{\pm 0.68}$ \\
      & JavaScript & 10,866 & 1,590 & 14.6$_{\pm 0.81}$ & 3.62 & 0.53 & 16.3$_{\pm 1.08}$ \\
      & Ruby & 10,290 & 2,904 & \textbf{28.2$_{\pm 2.29}$} & 3.43 & 0.97 & \textbf{26.9$_{\pm 0.70}$} \\
      & Rust & 10,908 & 2,022 & 18.5$_{\pm 0.29}$ & 3.64 & 0.67 & 17.0$_{\pm 0.85}$ \\ \cmidrule(lr){2-8}
      & \textit{Overall} & 41,958 & 7,716 & 18.4$_{\pm 0.56}$ & 10.49 & 1.93 & 18.1$_{\pm 0.17}$ \\
    \midrule
     \multirow{5}{*}{\texttt{llama-8b}}
      & Python & 17,016 & 4,176 & 24.5$_{\pm 0.95}$ & 11.34 & 2.78 & 20.4$_{\pm 0.26}$ \\
      & JavaScript & 17,796 & 4,362 & 24.5$_{\pm 0.40}$ & 11.86 & 2.91 & 23.0$_{\pm 0.62}$ \\
      & Ruby & 13,344 & 4,737 & 35.5$_{\pm 1.16}$ & 8.90 & 3.16 & \textbf{31.7$_{\pm 1.24}$} \\
      & Rust & 12,378 & 4,674 & \textbf{37.8$_{\pm 0.19}$} & 8.25 & 3.12 & 30.8$_{\pm 1.14}$ \\ \cmidrule(lr){2-8}
      & \textit{Overall} & 60,534 & 17,949 & 29.7$_{\pm 0.40}$ & 30.27 & 8.97 & 26.5$_{\pm 0.36}$ \\
    \midrule
     \multirow{5}{*}{\texttt{mistral-7b}}
      & Python & 4,821 & 597 & 12.4$_{\pm 0.38}$ & 3.21 & 0.40 & 9.1$_{\pm 0.14}$ \\
      & JavaScript & 4,713 & 360 & 7.6$_{\pm 0.42}$ & 3.14 & 0.24 & 7.4$_{\pm 0.37}$ \\
      & Ruby & 4,863 & 1,212 & \textbf{24.9$_{\pm 1.00}$} & 3.24 & 0.81 & \textbf{21.0$_{\pm 0.08}$} \\
      & Rust & 5,742 & 1,071 & 18.6$_{\pm 3.70}$ & 3.83 & 0.71 & 16.3$_{\pm 0.75}$ \\ \cmidrule(lr){2-8}
      & \textit{Overall} & 20,139 & 3,240 & 16.1$_{\pm 0.06}$ & 10.07 & 1.62 & 13.5$_{\pm 0.27}$ \\
    \midrule
     \multirow{5}{*}{\texttt{qwen-1.5b}}
      & Python & 23,553 & 8,280 & 35.1$_{\pm 1.60}$ & 7.85 & 2.76 & 21.1$_{\pm 0.84}$ \\
      & JavaScript & 28,320 & 11,454 & 40.4$_{\pm 3.33}$ & 9.44 & 3.82 & 23.8$_{\pm 0.98}$ \\
      & Ruby & 16,686 & 8,181 & 49.0$_{\pm 3.25}$ & 5.56 & 2.73 & 35.3$_{\pm 0.94}$ \\
      & Rust & 32,337 & 20,415 & \textbf{63.1$_{\pm 1.15}$} & 10.78 & 6.80 & \textbf{39.8$_{\pm 1.76}$} \\ \cmidrule(lr){2-8}
      & \textit{Overall} & 100,896 & 48,330 & 47.9$_{\pm 1.65}$ & 25.22 & 12.08 & 30.0$_{\pm 0.97}$ \\
    \midrule
     \multirow{5}{*}{\texttt{qwen-3b}}
      & Python & 6,864 & 1,458 & 21.2$_{\pm 2.36}$ & 2.29 & 0.49 & 20.6$_{\pm 0.23}$ \\
      & JavaScript & 6,348 & 1,662 & 26.2$_{\pm 2.28}$ & 2.12 & 0.55 & 25.4$_{\pm 0.52}$ \\
      & Ruby & 5,313 & 1,284 & 24.2$_{\pm 2.15}$ & 1.77 & 0.43 & 23.6$_{\pm 0.63}$ \\
      & Rust & 11,676 & 3,936 & \textbf{33.7$_{\pm 1.96}$} & 3.89 & 1.31 & \textbf{25.6$_{\pm 0.63}$} \\ \cmidrule(lr){2-8}
      & \textit{Overall} & 30,201 & 8,340 & 27.6$_{\pm 0.57}$ & 7.55 & 2.08 & 23.9$_{\pm 0.80}$ \\
    \bottomrule
    \end{tabular}
\end{table}

\begin{tcolorbox}[colback=gray!5, colframe=gray!50, title=\textbf{RQ1 Summary}]
Package hallucination is pervasive across all eight models and four languages
Rust and Ruby are challenging for the different studied models
and scaling from smaller to larger variants reduces hallucination by 16--21\,pp within each family. Yet no model is hallucination-free.
\end{tcolorbox}

\subsection{Impact of Guided Decoding Defenses}
\label{sec:rq2-strategies}

Here, we evaluate the effect of five guided decoding strategies (Greedy, Contrastive Decoding, DoLa, Nudging, and ALCD) and three baselines Vanilla decoding, RAG, and Self-Refine across all eight models and the four programming languages considered. Table~\ref{tab:strategy-macro-phr} shows the results. We observe that no single strategy dominates uniformly, indicating that hallucination mitigation is multi-factorial and strongly conditioned on both model family and language ecosystem.

\noindent
\textbf{RAG performs well for Python, Ruby, and Rust but degrades performance for JavaScript.}
RAG achieves the lowest PHR in 19 of 32 configurations and reduces the cross-model average from 29.7\% under Vanilla decoding to 18.1\%, the largest overall improvement among the evaluated strategies. It lowers PHR by an average of 10.7\,pp for Python, 23.4\,pp for Ruby, and 24.9\,pp for Rust, without increasing PHR for any model in these languages. In contrast, RAG increases PHR for 7 of 8 models on JavaScript, with an average degradation of 12.7\,pp. This suggests that the npm retrieval index or retrieval configuration requires careful validation before deployment.

\noindent
\textbf{DoLa performs poorly on DeepSeek but remains competitive for some model families.}
DoLa has the highest average PHR among the guided decoding strategies, reaching 38.0\% overall. Its performance deteriorates substantially on DeepSeek models, with DeepSeek-1.3b producing PHR values between 83.5\% and 91.2\%. However, DoLa remains moderately effective for models such as Mistral-7b and Qwen-3b, demonstrating that its effectiveness is architecture-dependent.

\noindent
\textbf{Self-Refine is most effective for Llama-8B.}
Self-Refine achieves the lowest PHR across all four languages for Llama-8B, reducing PHR by 8--18\,pp relative to Vanilla decoding. Its performance is less consistent for other models, where it frequently ranks among the weaker strategies. This suggests that successful self-correction depends on the capabilities of the underlying model rather than on architecture alone.

\noindent
\textbf{Nudging serves as the most dependable strategy for JavaScript.}
Where RAG performs poorly on JavaScript, Nudging consistently reduces PHR for Gemma-1b, Qwen-1.5b, Qwen-3b, and Llama-8b, making it the strongest strategy for this language overall. However, it degrades performance on DeepSeek models, further demonstrating that guided decoding strategies are not uniformly transferable across model families.

\noindent
\textbf{CD and ALCD provide limited gains, while Greedy decoding is a strong fallback.}
CD fails to achieve the lowest PHR in any configuration, while ALCD generally underperforms against Vanilla. In contrast, Greedy decoding matches Self-Refine's average PHR (25.6\%) and provides substantial reductions over Vanilla, making it a recommended fallback strategy.

\begin{table}[t]
\centering
\caption{Micro PHR (\%) per model, language, and strategy on the package recommendation task. \textbf{Bold} = lowest (best) per row.}
\label{tab:strategy-macro-phr}
\footnotesize
\setlength{\tabcolsep}{1pt}
\begin{tabular}{@{}llcccccccc@{}}
\toprule
\textbf{Model} & \textbf{Lang.} & \textbf{Vanilla} & \textbf{Greedy} & \textbf{Self-Refine} & \textbf{CD} & \textbf{DoLa} & \textbf{Nudge} & \textbf{RAG} & \textbf{ALCD} \\
\midrule
 \multirow{4}{*}{\texttt{deepseek-1.3b}}
  & Python & 35.6 & 44.3 & 36.2 & 44.1 & 87.3 & 42.7 & \textbf{18.3} & 66.3 \\
  & JavaScript & 30.0 & 38.8 & \textbf{20.0} & 38.5 & 83.5 & 42.8 & 41.5 & 70.0 \\
  & Ruby & 40.8 & 52.9 & 50.9 & 56.8 & 91.2 & 69.4 & \textbf{22.3} & 73.7 \\
  & Rust & 49.1 & 70.7 & 40.4 & 58.3 & 88.9 & 81.3 & \textbf{24.6} & 77.4 \\
\midrule
 \multirow{4}{*}{\texttt{deepseek-6.7b}}
  & Python & 12.6 & 12.6 & 26.4 & 46.6 & 79.2 & 33.1 & \textbf{9.4} & 41.4 \\
  & JavaScript & \textbf{10.2} & \textbf{10.2} & 15.6 & 40.6 & 88.8 & 32.0 & 36.0 & 46.6 \\
  & Ruby & 24.4 & 24.4 & 29.2 & 49.6 & 87.7 & 63.7 & \textbf{6.9} & 52.6 \\
  & Rust & 44.8 & 44.8 & 39.2 & 49.7 & 54.1 & 74.3 & \textbf{7.3} & 49.6 \\
\midrule
 \multirow{4}{*}{\texttt{gemma-1b}}
  & Python & 24.1 & 16.2 & 18.4 & 18.9 & 17.2 & 11.9 & \textbf{7.9} & 21.1 \\
  & JavaScript & 33.7 & \textbf{13.2} & 32.0 & 23.9 & 17.1 & 14.4 & 60.9 & 22.2 \\
  & Ruby & 53.9 & 33.4 & 30.0 & 37.1 & 39.3 & 30.2 & \textbf{3.5} & 40.3 \\
  & Rust & 39.9 & 31.1 & 36.7 & 38.9 & 35.3 & 24.8 & \textbf{2.9} & 36.7 \\
\midrule
 \multirow{4}{*}{\texttt{gemma-4b}}
  & Python & 12.1 & 12.5 & 10.8 & 11.6 & 16.9 & 11.9 & \textbf{6.7} & 16.7 \\
  & JavaScript & 14.6 & 12.5 & 14.4 & 15.0 & 20.3 & \textbf{11.7} & 12.0 & 14.6 \\
  & Ruby & 28.2 & 24.4 & 23.6 & 20.1 & 42.4 & 23.9 & \textbf{8.7} & 29.2 \\
  & Rust & 18.5 & 17.8 & 18.6 & 19.2 & 22.7 & 14.5 & \textbf{8.6} & 17.0 \\
\midrule
 \multirow{4}{*}{\texttt{llama-8b}}
  & Python & 24.5 & 24.5 & \textbf{10.8} & 21.6 & 20.9 & 11.4 & 18.9 & 21.9 \\
  & JavaScript & 24.5 & 24.5 & \textbf{9.9} & 18.2 & 22.3 & 12.7 & 28.6 & 22.4 \\
  & Ruby & 35.5 & 35.5 & \textbf{17.6} & 24.6 & 36.2 & 19.0 & 22.7 & 35.6 \\
  & Rust & 37.8 & 37.8 & \textbf{27.3} & 33.5 & 33.4 & 28.9 & 28.7 & 34.1 \\
\midrule
 \multirow{4}{*}{\texttt{mistral-7b}}
  & Python & 12.4 & 14.6 & 10.3 & 13.4 & 10.1 & 11.7 & 11.3 & \textbf{9.9} \\
  & JavaScript & 7.6 & 7.6 & 9.4 & 12.3 & \textbf{7.0} & 10.4 & 19.9 & 7.6 \\
  & Ruby & 24.9 & 24.6 & 25.4 & 18.4 & 20.2 & 17.2 & \textbf{16.5} & 20.3 \\
  & Rust & 18.6 & 23.4 & 20.0 & 17.2 & 15.9 & 19.2 & 20.8 & \textbf{14.4} \\
\midrule
 \multirow{4}{*}{\texttt{qwen-1.5b}}
  & Python & 35.1 & 14.9 & 16.9 & 24.2 & 13.2 & 15.7 & \textbf{7.7} & 28.8 \\
  & JavaScript & 40.4 & 18.0 & 22.6 & 28.7 & 16.8 & \textbf{12.3} & 47.5 & 28.6 \\
  & Ruby & 49.0 & 28.5 & 46.5 & 35.6 & 32.1 & 20.5 & \textbf{7.4} & 40.1 \\
  & Rust & 63.1 & 31.7 & 48.2 & 46.3 & 34.3 & 23.8 & \textbf{5.1} & 55.4 \\
\midrule
 \multirow{4}{*}{\texttt{qwen-3b}}
  & Python & 21.2 & 15.3 & 22.3 & 14.9 & 13.5 & 12.5 & \textbf{11.5} & 15.0 \\
  & JavaScript & 26.2 & 21.1 & 20.9 & 20.8 & 24.4 & \textbf{12.0} & 42.5 & 22.7 \\
  & Ruby & 24.2 & 18.8 & 32.1 & 25.4 & 23.2 & 20.4 & \textbf{5.7} & 22.4 \\
  & Rust & 33.7 & 19.9 & 36.8 & 28.9 & 20.3 & 18.9 & \textbf{8.1} & 27.3 \\
\bottomrule
\end{tabular}
\end{table}

\subsection{Extension to Full Code Generation, Syntax Validity and Code Smells}
\label{sec:full-code-generation}
In our previous experiments in Sections \ref{sec:library-correction}, \ref{sec:rq1-baseline}, and \ref{sec:rq2-strategies}, we prompted each LLM to recommend a set of useful packages for every instruction in $\mathcal{D}_{\text{eval}}$. We extend this analysis to full code generation by prompting each model to produce complete, runnable code for the same instructions. All imported packages are validated against the same registry snapshots.
The system prompt for this task was:
\begin{tcolorbox}[title={System Prompt: Code Generation},
fonttitle=\footnotesize,
fontupper=\footnotesize,
boxsep=1pt,
    left=2mm,
    right=2mm,
    top=1mm,
    bottom=1mm,
    toptitle=1mm,
    bottomtitle=1mm,
    before skip=2pt,
    after skip=2pt]
    You are a helpful assistant that generates \texttt{<language>} code for a given task.
    Respond with code inside a \texttt{```<language> ... ```} block when appropriate.
\end{tcolorbox}

\begin{figure*}[t!]
    \centering
    \begin{subfigure}[t]{0.49\textwidth}
        \centering
        \includegraphics[width=0.98\linewidth]{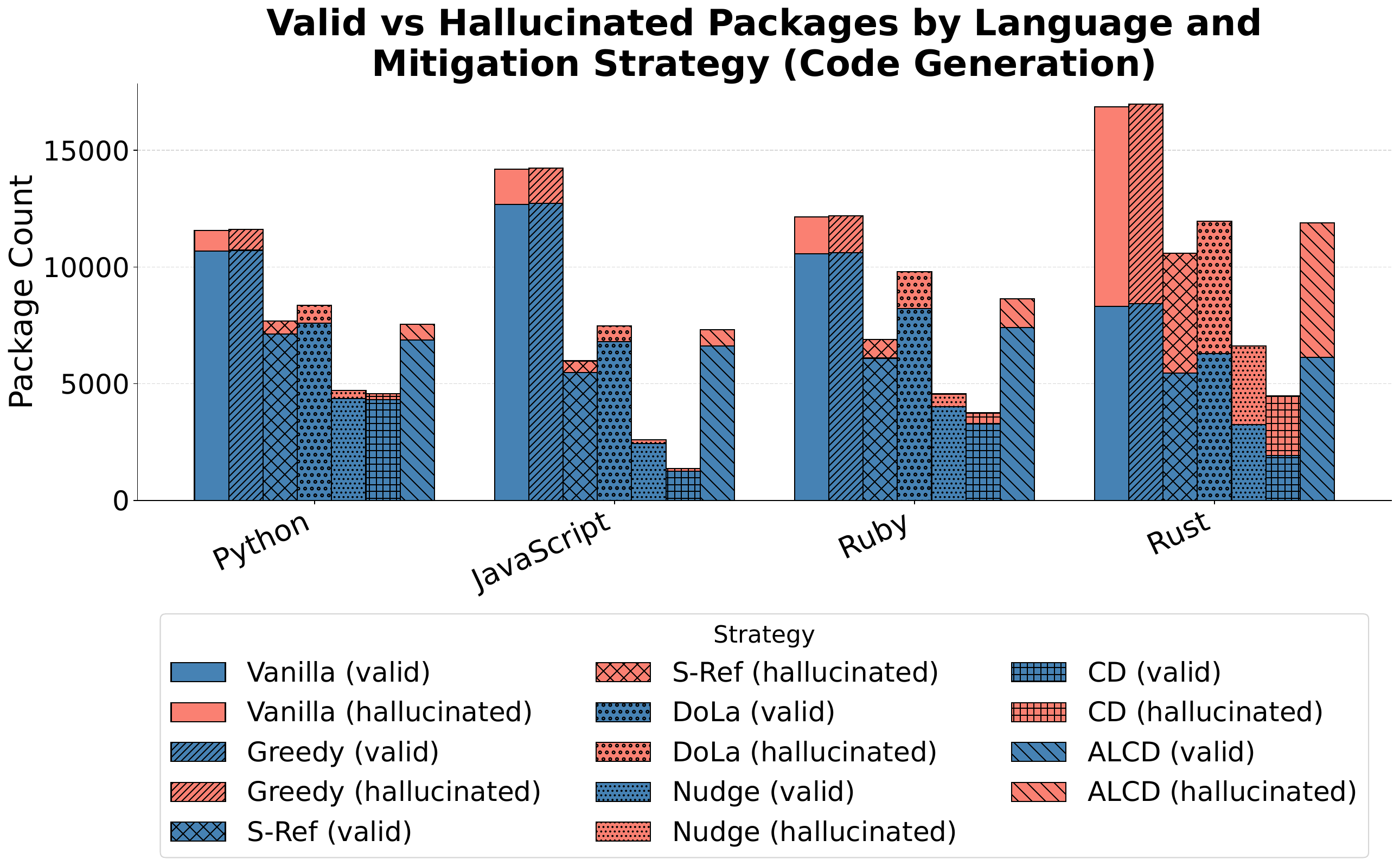}
        \caption{Distribution of valid vs. hallucinated packages by language and strategy for full code generation.}
        \label{fig:dist-lang-method-code}
    \end{subfigure}
    \hfill
    \begin{subfigure}[t]{0.49\textwidth}
       \centering
        \includegraphics[width=0.98\linewidth]{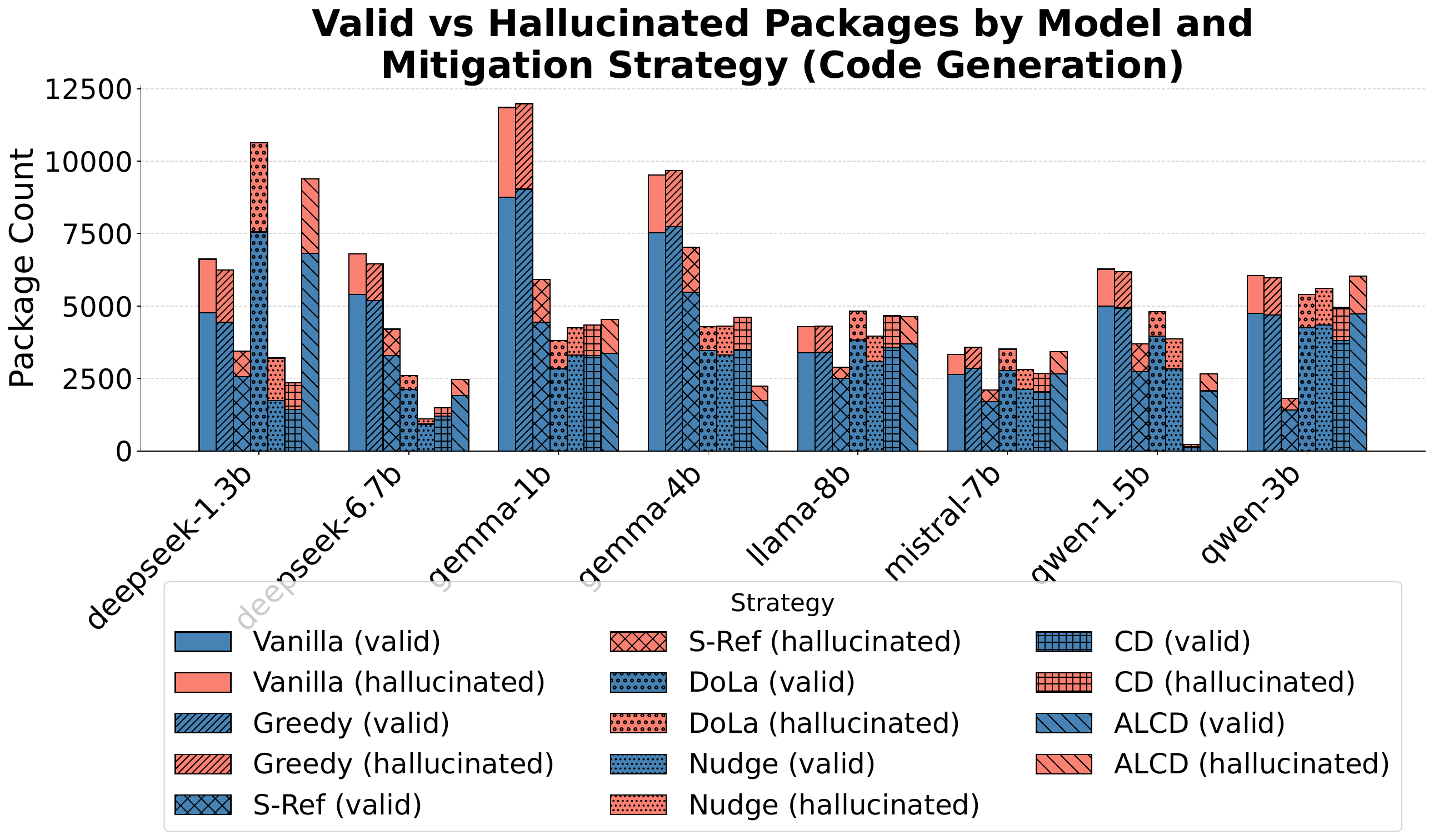}
        \caption{Distribution of valid vs. hallucinated packages by model and strategy for full code generation.}
        \label{fig:dist-model-method-code}
    \end{subfigure}
    \caption{Valid and hallucinated package references extracted from full code generation outputs, aggregated by language (left) and by model (right) across all seven strategies. Blue segments denote registry-validated packages; red segments denote hallucinated references.}
    \Description{Bar plots showing the distribution of valid and hallucinated package references for code generation outputs, separated by language and model. Blue bars represent valid packages, red bars represent hallucinated packages.}
\end{figure*}

We exclude RAG from this experiment because, in our setup, it affects only package recommendation and does not modify the subsequent code-generation process. Figures~\ref{fig:dist-lang-method-code} and~\ref{fig:dist-model-method-code} show the results of valid and hallucinated packages in generated code by language and model. At the model level (Figure~\ref{fig:dist-model-method-code}), Self-Refine produces the most package references but increases both valid and hallucinated counts, indicating that its gains in package recommendation do not transfer reliably to code generation. Mistral-7b produces the fewest references while maintaining the most favorable valid-to-hallucinated ratio, reflecting more conservative dependency use. Nudging preserves more valid references than Vanilla across models, whereas CD reduces both valid and hallucinated references, suggesting output suppression. At the language level (Figure~\ref{fig:dist-lang-method-code}), Rust exhibits the highest hallucination volume, nearly matching its valid-reference count. For JavaScript, Self-Refine substantially increases hallucinated references, while DoLa produces very few references overall. Python is the most controlled setting, with Nudging maintaining a clearer separation between valid and hallucinated references.

\begin{figure*}[t!]
    \centering
    \begin{subfigure}[t]{0.49\textwidth}
       \centering
        \includegraphics[width=0.99\linewidth]{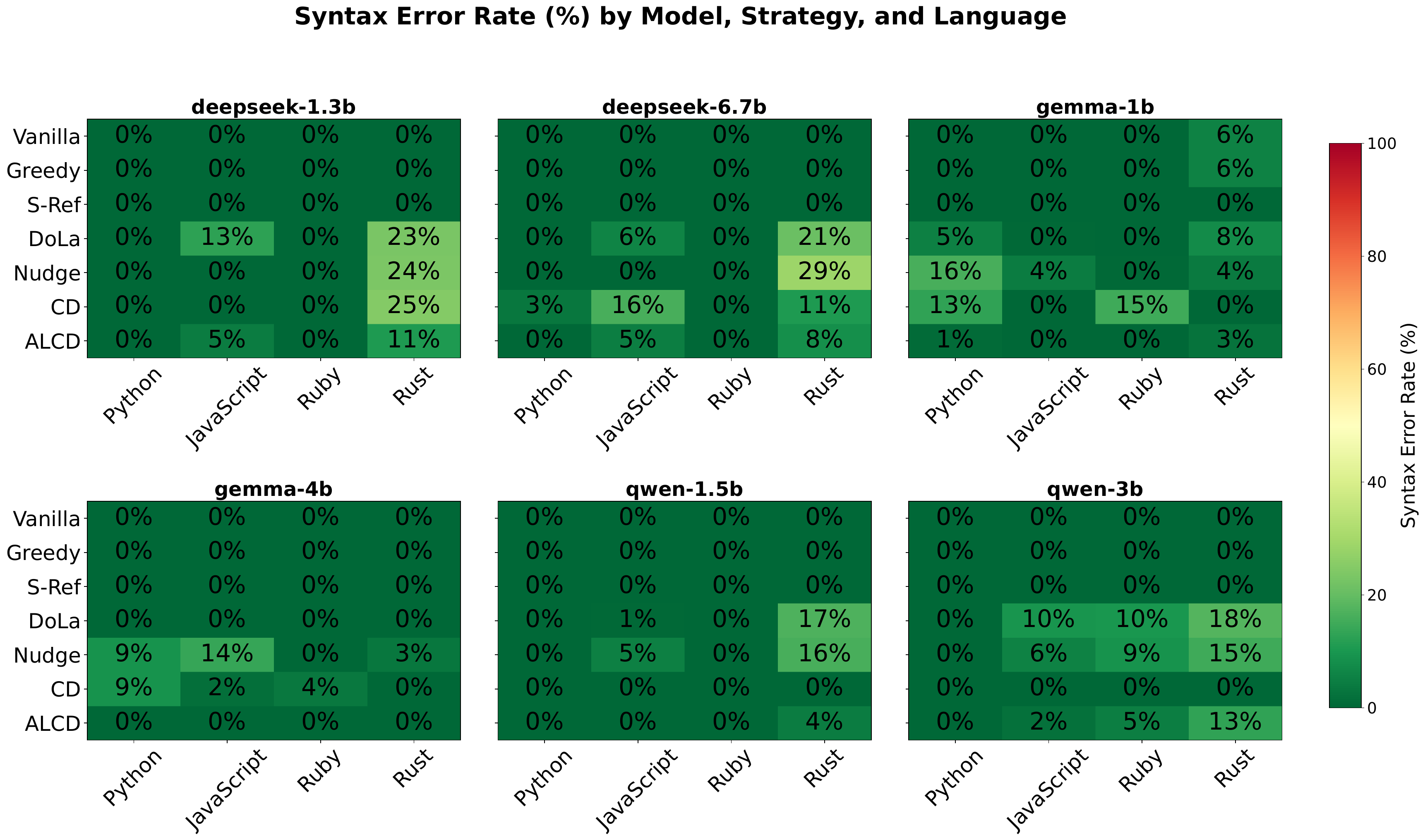}
        \Description{Heatmap showing syntax error rates per model, strategy, and language.}
        \caption{Syntax error rate (\%) per model, strategy, and language.}
        \label{fig:syntax_heatmap}
    \end{subfigure}
    \hfill
    \begin{subfigure}[t]{0.49\textwidth}
       \centering
        \includegraphics[width=0.99\linewidth]{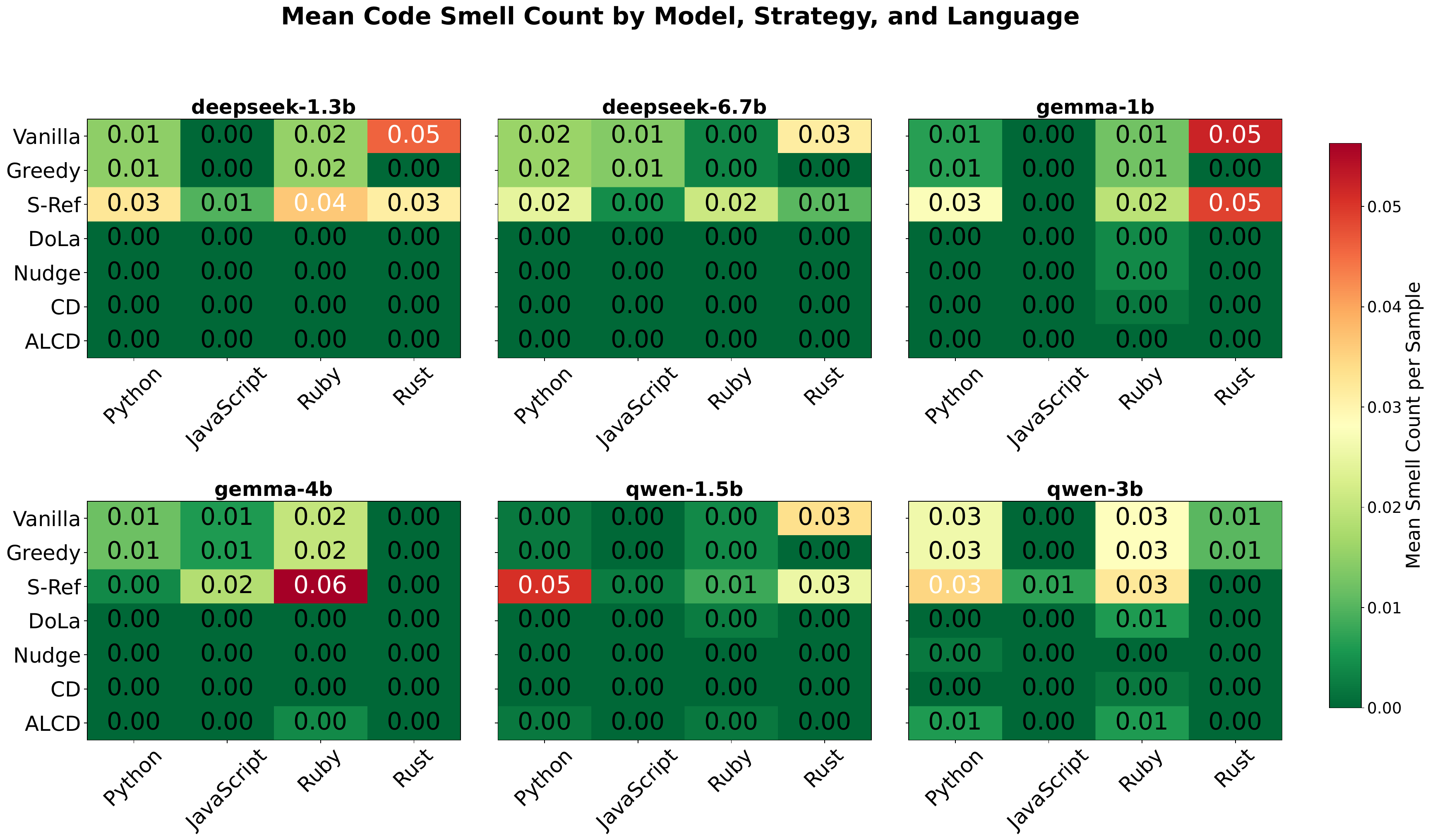}
        \caption{Code smell rate (\%) per model, strategy, and language.}
        \label{fig:smell_heatmap}
    \end{subfigure}
    \caption{Structural quality of generated code per model, strategy, and language. \textbf{Left:} syntax error rate (\%); green cells denote error-free output, warmer colors indicate increasing parse failure rates. \textbf{Right:} mean Semgrep-detected code smell count per sample.}
\end{figure*}

\noindent
\textbf{Syntax Validity and Code Smells.} Beyond package hallucination, we assess the structural quality of generated code using Semgrep, measuring two complementary metrics: \ding{182} \emph{syntax error rate} (fraction of programs that could not be parsed due to a syntax error) and \ding{183} \emph{code smell rate} (fraction exhibiting Semgrep-detected anti-patterns). Results are reported in Figure~\ref{fig:syntax_heatmap} and~\ref{fig:smell_heatmap}.
We observe that Vanilla and Greedy decoding ensure nearly perfect syntactic validity across six models, while Self-Refine maintains clean syntax despite being the least effective for code smells. Guided strategies, however, consistently introduce syntax errors, particularly in Rust and JavaScript. Notably, \emph{Nudging} exhibits the most heterogeneous syntax degradation, with DeepSeek-6.7b showing a 29\% syntax error rate for Rust. In contrast, Gemma-4b stands out by producing 0\% syntax errors across all languages analyzed.
For code smells, DoLa, Nudging, and CD yield nearly zero smell counts across most model-language combinations, whereas Vanilla and Self-Refine are significant sources of anti-patterns. Vanilla decoding sees the highest smells in Rust models, while Self-Refine exacerbates these issues, particularly in Gemma-4b for Ruby. Overall, these findings suggest that certain strategies can lead to both syntax errors and code smells, highlighting the importance of the modeling approach used.

\begin{tcolorbox}[colback=gray!5, colframe=gray!50, title=\textbf{RQ2 Summary}]
No single strategy dominates across models, languages, and tasks. RAG achieves the lowest PHR in 19/32 package-recommendation configurations for Python, Ruby, and Rust but degrades JavaScript by $12.7$\,pp on average; Self-Refine is strongest for Llama-8B. In code generation, Vanilla and Greedy preserve syntax, whereas guided strategies can introduce parsing errors but often reduce code smells. Strategy selection must therefore consider the model family, language, task, and broader code quality.

\end{tcolorbox}

\subsection{Package Utility of Guided Decoding}
\label{sec:rq3-du}

\begin{figure*}[t!]
    \centering
    \begin{subfigure}[t]{0.98\textwidth}
        \centering
        \includegraphics[width=\linewidth]{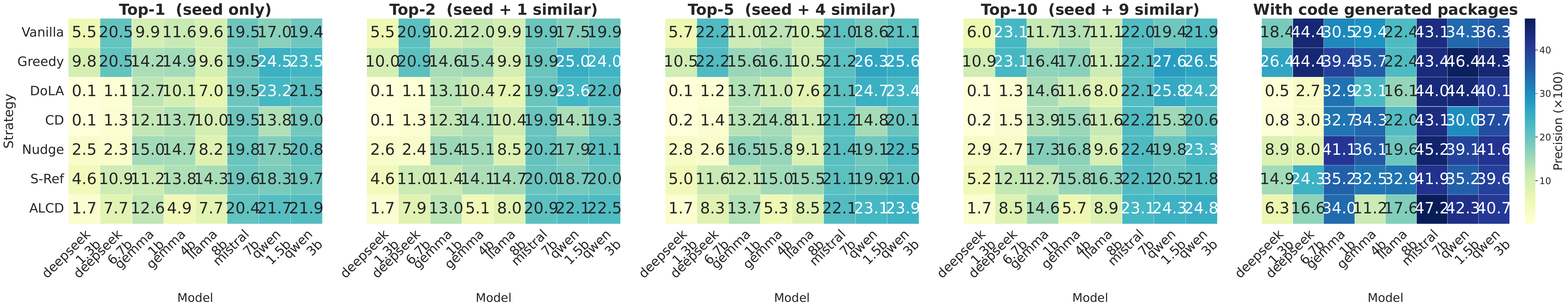}
        \caption{Package Utility Precision $\mathsf{PU_P}$ ($\times 100$).}
        \label{fig:du_precision}
    \end{subfigure}
    \vspace{4pt}
    \begin{subfigure}[t]{0.98\textwidth}
        \centering
        \includegraphics[width=\linewidth]{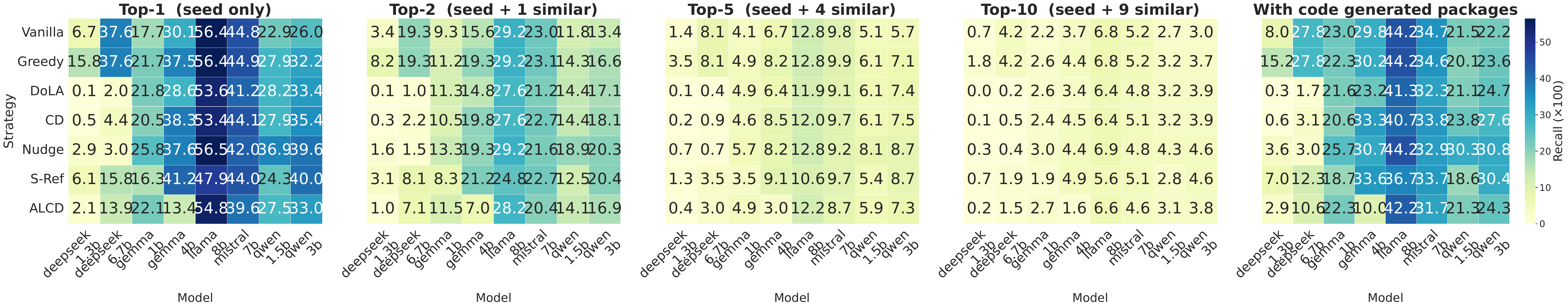}
        \caption{Package Utility Recall $\mathsf{PU_R}$ ($\times 100$).}
        \label{fig:du_recall}
    \end{subfigure}
    \vspace{4pt}
    \begin{subfigure}[t]{0.98\textwidth}
        \centering
        \includegraphics[width=\linewidth]{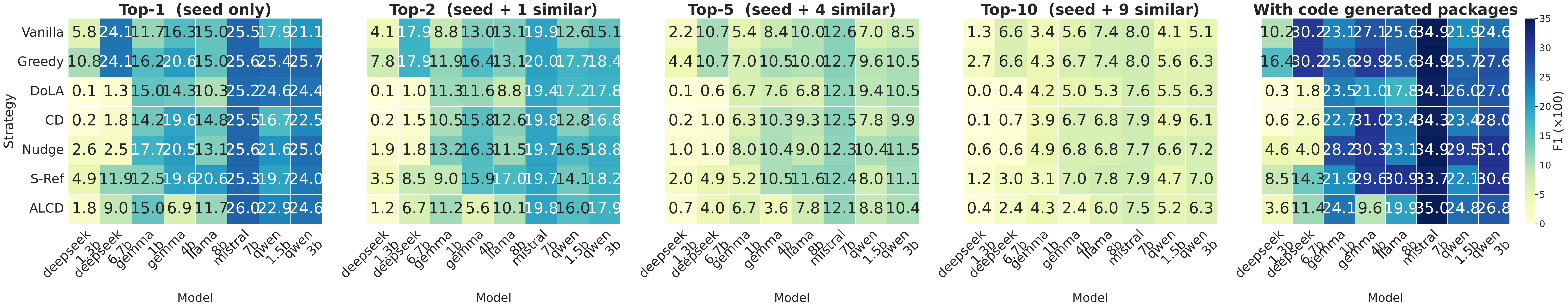}
        \caption{Package Utility $\mathsf{PU}$ ($\times 100$).}
        \label{fig:du_f1}
    \end{subfigure}
    \caption{Package Utility ($\mathsf{PU}$) across five reference-set definitions for all seven strategies and eight models on the package recommendation task. Values are multiplied by 100 for readability. Rows: strategies; columns: models.}
    \label{fig:du_topk}
\end{figure*}

We assess whether model-recommended packages are useful for the target task, rather than merely valid. We measure this using Package Utility ($\mathsf{PU}$) (see Section \ref{sec:package-utility}), a precision--recall metric computed against a task-specific reference set $R$. We consider two reference-set constructions: \ding{182} the top-$k$ packages whose description is most similar to the seed package in terms of cosine similarity, where $k \in \{1,2,5,10\}$, and \ding{183} a code-based reference set containing packages obtained in the code generation experiment in Section \ref{sec:full-code-generation}. Figures~\ref{fig:du_precision}--\ref{fig:du_f1} report the micro-averaged precision ($\mathsf{PU_P}$), recall ($\mathsf{PU_R}$), and overall ($\mathsf{PU}$) scores.

\noindent
\textbf{Precision is insensitive to \( k \) while recall is not.} 
Across strategies and models, micro $\mathsf{PU_P}$ increases by marginally 1–2 percentage points from Top-1 to Top-10 (mean values from 13.1 to 14.8), despite the tenfold growth of the reference set. This suggests that models recommend a limited set of packages based on training distribution rather than by valid reference coverage, indicating that precision reflects generation selectivity. In contrast, micro $\mathsf{PU_R}$ decreases from a mean of 23.4 at Top-1 to only 2.7 at Top-10, an 8.6-fold reduction. This trend indicates that models capture a narrow scope of semantically proximate packages, with no strategy achieving a mean recall above 8 at Top-10.

\noindent
\textbf{Reference based on generated code packages set enhances utility recovery.} 
The generated code package reference results in significantly higher metric values, with mean $\mathsf{PU}$ rising to 21.0 from 4.3 at Top-10 and 15.5 at Top-1. This improvement reflects the alignment between the packages obtained from the generated code given a prompt sets and the recommended packages list provided by the model given the same prompt, while establishing that this approach provide informative operational definition of $\mathsf{PU}$.

\noindent
\textbf{DeepSeek utility collapse confirmed across all definitions.} 
Both DeepSeek variants show near-zero $\mathsf{PU}$ under all tested conditions, reinforcing previous findings of output suppression. Conversely, Baseline and Greedy on DeepSeek-6.7b achieve a collective $\mathsf{PU}$ of 30.2, indicating retained latent utility.

\noindent
\textbf{Mitigation--utility trade-off.}
Among strategies evaluated on both PHR and $\mathsf{PU}$, Greedy provides the strongest average trade-off, reducing mean PHR from 29.7\% under Vanilla to 25.6\% while attaining the highest $\mathsf{PU}$ under all five reference-set definitions. Self-Refine ranks second followed by Nudging. CD, DoLa, and ALCD do not reduce average PHR relative to Vanilla and therefore provide an unfavorable overall trade-off.

\begin{tcolorbox}[colback=gray!5, colframe=gray!50, title=\textbf{RQ3 Summary}]

Package utility depends strongly on reference-set construction. Expanding the similarity-based set from Top-1 to Top-10 leaves precision nearly unchanged but reduces recall drastically from 23.4 to 2.7 and mean $\mathsf{PU}$ from 15.5 to 4.3. Code-derived references yield the highest mean $\mathsf{PU}$, providing a more operational measure of utility.
\end{tcolorbox}

\subsection{Adversarial Robustness}
\label{sec:rq4-adversarial}

A defense that works only under benign instructions offers limited practical
assurance \cite{euraste2026learned, djire2025memorization, olatunji2025adversarial, rauch2021achieving}. In this section we evaluate all mitigation strategies under adversarial
prompts where the user explicitly directs the model to install and use fabricated
packages.

\noindent
\textbf{Adversarial Prompt Construction.}
We collected 1,000 non-existent package names for each Python and JavaScript language produced by both open-source
(Meta-Llama, Alibaba families) and closed-source (OpenAI GPT family) LLMs when
prompted on coding tasks. Each candidate was verified as absent from the npm and
PyPI registries via direct lookup.
These names cover a broad
spectrum of hallucination patterns including completely invented identifiers,
cross-ecosystem confusions, fuzzy variants of real names, standard-library
misattributions, and case or naming-convention mismatches.
To extend coverage to Ruby and Rust, we reused these npm- and PyPI-origin names as
adversarial seeds. This choice is deliberate: cross-ecosystem confusion is itself
one of the most prevalent hallucination patterns, and a model asked to use a
JavaScript-origin name in a Ruby context may hallucinate a plausible gem rather
than refuse. Each name was inserted into a language-specific instruction template
directing a coding assistant to install and use the corresponding package, yielding
1,000 adversarial prompts per language (4,000 total).

\noindent
\textbf{Impact of Adversarial Prompting.}
Table~\ref{tab:adversarial-phr} shows that adversarial prompts substantially increase PHR relative to the vanilla baselines. Averaged across the four languages, we observe an increment ranging from 20.6\,pp
The effect is particularly severe for models that perform well under benign prompts. Gemma-4b, whose vanilla PHR ranges from 12.1\% to 28.2\%, reaches 49.7\%--90.2\% under attack. Similarly, Mistral-7b, the strongest model under vanilla decoding (7.6\%--24.9\%), rises to 46.6\%--76.6\%, an average increase of 47.8\,pp. Thus, strong performance under benign prompts provides limited protection once the prompt itself is adversarially framed.
Qwen-1.5b shows the smallest increase. This can be attributed to its vanilla PHR already being the highest in
the study (35.1--63.1\%); on Rust, its adversarial PHR (51.4\%) is in fact
lower than its vanilla rate (63.1\%), the single case where adversarial framing does not increase hallucination.
Ruby is the most vulnerable language and Rust the most resistant in each of the eight models we test, with no exception. Adversarial PHR under vanilla decoding spans 46.6\% (Mistral-7b, Rust) to 92.5\%
(DeepSeek-1.3b, Ruby). This universality suggests that Ruby's smaller
package ecosystem provides weaker negative-evidence anchors in pre-training corpora
regardless of model family or scale, while Rust's strict compiler-enforced naming
conventions consistently constrain fabrication.

\noindent
\textbf{Effectiveness of Mitigation Strategies.}
Table~\ref{tab:adversarial-phr} shows that no single strategy dominates across all eight models, although a scale-dependent pattern emerges. RAG performs best overall, achieving the lowest PHR in 16 of 32 model-language combinations (50\%). It is particularly effective for lower-capacity models, producing the lowest PHR on 3 of 4 languages for DeepSeek-1.3b (Python: 48.6\%, JavaScript: 71.8\%, Ruby: 87.2\%) and Qwen-1.5b (Python: 24.5\%, JavaScript: 46.9\%, Ruby: 64.2\%), on all 4 languages for DeepSeek-6.7b, and on 2 of 4 for Gemma-1B (Python: 39.0\%, JavaScript: 49.2\%). Retrieved registry evidence therefore appears especially useful for models with limited self-correction capacity. The JavaScript degradation observed under benign prompts (Section~\ref{sec:rq2-strategies}) is absent under attack, where RAG achieves the lowest JavaScript PHR for 5 of 8 models.
Self-Refine performs more consistently on models with stronger self-correction capabilities, yielding the lowest PHR in 13 of 32 combinations. It ranks first across all four languages for Llama-8B, three of four for Mistral-7b, and for Gemma-4b on JavaScript (31.0\%), Ruby (69.8\%), and Rust (38.6\%). It also performs best for Qwen-3b on Ruby (57.9\%) and Rust (44.3\%). The lowest PHR observed in the study is 11.4\%, achieved by Gemma-4b on Python with RAG. This result indicates that retrieval grounding can provide strong resistance to adversarial package injection even for a 4b model. Together, RAG and Self-Refine produce the lowest PHR in 29 of 32 combinations. Nudging performs best for DeepSeek-1.3b and Gemma-1b on Rust. For Gemma-1B on Ruby, no mitigation strategy improves upon the baseline PHR of 83.7\%. This result shows that, for small models operating in under-represented package ecosystems, the evaluated interventions may fail to reduce hallucination and can instead increase it.

\noindent
\textbf{Decoding-only strategies provide inconsistent protection under adversarial prompting.}
DoLa increases PHR relative to Greedy in 24 of 32 model--language configurations (75.0\%). The largest degradation occurs for DeepSeek-1.3B on Rust, where PHR rises from 38.6\% under Greedy to 94.1\% under DoLa, an increase of 55.5\,pp. Similar failures occur for Gemma-4b on Rust, with PHR increasing from 49.5\% to 87.1\% ($+37.6$\,pp), and for Mistral-7b on Rust, where it increases from 46.1\% to 81.3\% ($+35.2$\,pp). DoLa underperforms Greedy across all four languages for Gemma-1b, Qwen-3b, Llama-8b, and Mistral-7b. Most improvements are concentrated in the DeepSeek family: both DeepSeek-1.3b and DeepSeek-6.7b outperform Greedy in three of four languages, suggesting that DoLa's layer selection may be better aligned with DeepSeek's internal representations. ALCD mitigates some of these failures by applying layer contrast selectively. It outperforms DoLa in 23 of 32 configurations, including reductions from 81.6\% to 51.0\% for Llama-8b on Rust and from 49.8\% to 36.5\% for Qwen-1.5b on Rust. ALCD also improves upon DoLa across all four languages for DeepSeek-6.7b, Qwen-1.5b, and Qwen-3b. However, it does not achieve the lowest PHR in any configuration and improves upon the unmitigated Vanilla baseline in only 6 of 32 cases, all involving DeepSeek-1.3b, Gemma-4B, or Qwen-1.5b on Rust. Thus, ALCD reduces DoLa's degradation but remains an unreliable adversarial defense.
Nudging and CD also produce mixed results. Nudging improves upon Greedy across all four languages for DeepSeek-1.3b, including a reduction from 38.6\% to 26.6\% on Rust ($-12.0$\,pp). It also reduces PHR for Gemma-1b on Rust from 75.0\% to 47.6\% ($-27.4$\,pp). In contrast, PHR increases from 63.1\% to 72.5\% for Qwen-1.5b on JavaScript ($+9.4$\,pp) and from 76.8\% to 84.8\% for Gemma-1b on JavaScript ($+8.0$\,pp). 
These results confirm that Nudging remains model- and language-dependent. CD does not achieve the lowest PHR in any configuration and consistently trails RAG and Self-Refine. Greedy remains close to the Vanilla baseline in most cases, indicating that deterministic decoding alone provides limited adversarial robustness.

\noindent
\textbf{Why decoding-only defenses fail under adversarial prompts.} 
Decoding-only methods operate on local token probabilities or internal representations, but they do not verify whether a package exists. When the prompt contains a fabricated package, the fake name becomes a strong contextual anchor. Without external registry evidence or explicit self-verification, guided decoding may preserve or even amplify plausible-looking package names rather than reject them. This explains why RAG and Self-Refine are more robust under adversarial prompts. Specifically, RAG introduces external registry grounding, while Self-Refine gives the model an explicit opportunity to reconsider generated dependencies.

\begin{table*}[t]
    \centering
    \caption{Micro PHR (\%) and raw counts ($N_{\text{hall}}$/$N_{\text{gen}}$) in parentheses per model, language, and strategy under adversarial prompts. Lowest PHR per row in \textbf{bold}.}
    \label{tab:adversarial-phr}
    \footnotesize
    \setlength{\tabcolsep}{5pt}
    \begin{tabular}{@{}llcccccccc@{}}
    \toprule
    \textbf{Model} & \textbf{Lang.} &
    \textbf{Vanilla} & \textbf{Greedy} & \textbf{Self-Refine} &
    \textbf{DoLa} & \textbf{RAG} & \textbf{Nudge} & \textbf{CD} &
    \textbf{ALCD}
  \\
    \midrule

    \multirow{4}{*}{\texttt{deepseek-1.3b}}
      & JavaScript
        & \makecell{84.4  ({\tiny 320/379})}
        & \makecell{87.4  ({\tiny 340/389})}
        & \makecell{75.2  ({\tiny 194/258})}
        & \makecell{84.9  ({\tiny 253/298})}
        & \makecell{\textbf{71.8}  ({\tiny 492/685})}
        & \makecell{76.9  ({\tiny 230/299})}
        & \makecell{87.7  ({\tiny 193/220})}
        & \makecell{76.1  ({\tiny 242/318})} \\
      & Python
        & \makecell{80.3  ({\tiny 392/488})}
        & \makecell{90.4  ({\tiny 339/375})}
        & \makecell{81.2  ({\tiny 234/288})}
        & \makecell{68.5  ({\tiny 204/298})}
        & \makecell{\textbf{48.6}  ({\tiny 470/968})}
        & \makecell{80.0  ({\tiny 419/524})}
        & \makecell{93.9  ({\tiny 845/900})}
        & \makecell{73.6  ({\tiny 245/333})} \\
      & Ruby
        & \makecell{92.5  ({\tiny 368/398})}
        & \makecell{93.2  ({\tiny 449/482})}
        & \makecell{91.8  ({\tiny 180/196})}
        & \makecell{93.1  ({\tiny 244/262})}
        & \makecell{\textbf{87.2}  ({\tiny 429/492})}
        & \makecell{91.2  ({\tiny 466/511})}
        & \makecell{96.8  ({\tiny 987/1{,}020})}
        & \makecell{87.3  ({\tiny 207/237})} \\
      & Rust
        & \makecell{72.4  ({\tiny 422/583})}
        & \makecell{38.6  ({\tiny 248/643})}
        & \makecell{91.2  ({\tiny 312/342})}
        & \makecell{94.1  ({\tiny 636/676})}
        & \makecell{79.3  ({\tiny 466/588})}
        & \makecell{\textbf{26.6}  ({\tiny 238/895})}
        & \makecell{91.9  ({\tiny 570/620})}
        & \makecell{86.0  ({\tiny 301/350})} \\
    \midrule

    \multirow{4}{*}{\texttt{deepseek-6.7b}}
      & JavaScript
        & \makecell{73.1  ({\tiny 425/581})}
        & \makecell{81.3  ({\tiny 231/284})}
        & \makecell{86.0  ({\tiny 505/587})}
        & \makecell{91.6  ({\tiny 164/179})}
        & \makecell{\textbf{61.8}  ({\tiny 777/1{,}257})}
        & \makecell{96.7  ({\tiny 205/212})}
        & \makecell{76.4  ({\tiny 292/382})}
        & \makecell{86.1  ({\tiny 167/194})} \\
      & Python
        & \makecell{73.2  ({\tiny 423/578})}
        & \makecell{83.6  ({\tiny 316/378})}
        & \makecell{59.5  ({\tiny 601/1{,}010})}
        & \makecell{81.0  ({\tiny 111/137})}
        & \makecell{\textbf{41.0}  ({\tiny 1{,}156/2{,}819})}
        & \makecell{88.1  ({\tiny 244/277})}
        & \makecell{60.1  ({\tiny 196/326})}
        & \makecell{75.0  ({\tiny 165/220})} \\
      & Ruby
        & \makecell{86.1  ({\tiny 353/410})}
        & \makecell{95.2  ({\tiny 479/503})}
        & \makecell{87.8  ({\tiny 360/410})}
        & \makecell{90.0  ({\tiny 199/221})}
        & \makecell{\textbf{77.5}  ({\tiny 1{,}418/1{,}829})}
        & \makecell{95.7  ({\tiny 485/507})}
        & \makecell{83.4  ({\tiny 151/181})}
        & \makecell{89.9  ({\tiny 249/277})} \\
      & Rust
        & \makecell{66.9  ({\tiny 521/779})}
        & \makecell{85.2  ({\tiny 574/674})}
        & \makecell{81.1  ({\tiny 439/541})}
        & \makecell{74.3  ({\tiny 182/245})}
        & \makecell{\textbf{58.9}  ({\tiny 1{,}328/2{,}253})}
        & \makecell{94.3  ({\tiny 528/560})}
        & \makecell{96.3  ({\tiny 1{,}423/1{,}478})}
        & \makecell{67.0  ({\tiny 185/276})} \\
    \midrule

    \multirow{4}{*}{\texttt{gemma-1b}}
      & JavaScript
        & \makecell{77.0  ({\tiny 1{,}151/1{,}494})}
        & \makecell{76.8  ({\tiny 994/1{,}295})}
        & \makecell{78.5  ({\tiny 707/901})}
        & \makecell{88.5  ({\tiny 571/645})}
        & \makecell{\textbf{49.2}  ({\tiny 883/1{,}793})}
        & \makecell{84.8  ({\tiny 1{,}180/1{,}391})}
        & \makecell{77.6  ({\tiny 1{,}303/1{,}680})}
        & \makecell{88.6  ({\tiny 658/743})} \\
      & Python
        & \makecell{72.6  ({\tiny 1{,}224/1{,}685})}
        & \makecell{74.8  ({\tiny 1{,}265/1{,}691})}
        & \makecell{66.7  ({\tiny 873/1{,}309})}
        & \makecell{84.6  ({\tiny 737/871})}
        & \makecell{\textbf{39.0}  ({\tiny 640/1{,}640})}
        & \makecell{73.4  ({\tiny 1{,}144/1{,}558})}
        & \makecell{75.1  ({\tiny 1{,}383/1{,}841})}
        & \makecell{79.6  ({\tiny 930/1{,}168})} \\
      & Ruby
        & \makecell{\textbf{83.7}  ({\tiny 1{,}113/1{,}329})}
        & \makecell{86.3  ({\tiny 1{,}073/1{,}244})}
        & \makecell{88.4  ({\tiny 729/825})}
        & \makecell{86.4  ({\tiny 654/757})}
        & \makecell{86.1  ({\tiny 802/931})}
        & \makecell{87.4  ({\tiny 587/672})}
        & \makecell{86.3  ({\tiny 1{,}116/1{,}293})}
        & \makecell{86.6  ({\tiny 775/895})} \\
      & Rust
        & \makecell{69.2  ({\tiny 1{,}162/1{,}679})}
        & \makecell{75.0  ({\tiny 1{,}134/1{,}511})}
        & \makecell{75.0  ({\tiny 496/661})}
        & \makecell{88.2  ({\tiny 464/526})}
        & \makecell{69.0  ({\tiny 459/665})}
        & \makecell{\textbf{47.6}  ({\tiny 1{,}068/2{,}243})}
        & \makecell{48.0  ({\tiny 828/1{,}726})}
        & \makecell{90.9  ({\tiny 680/748})} \\
    \midrule

    \multirow{4}{*}{\texttt{gemma-4b}}
      & JavaScript
        & \makecell{88.6  ({\tiny 1{,}692/1{,}909})}
        & \makecell{89.4  ({\tiny 1{,}774/1{,}984})}
        & \makecell{\textbf{31.0}  ({\tiny 567/1{,}828})}
        & \makecell{82.6  ({\tiny 804/973})}
        & \makecell{71.1  ({\tiny 927/1{,}303})}
        & \makecell{88.1  ({\tiny 1{,}667/1{,}892})}
        & \makecell{86.2  ({\tiny 1{,}692/1{,}963})}
        & \makecell{84.1  ({\tiny 630/749})} \\
      & Python
        & \makecell{77.2  ({\tiny 1{,}232/1{,}596})}
        & \makecell{79.1  ({\tiny 1{,}234/1{,}560})}
        & \makecell{37.7  ({\tiny 590/1{,}563})}
        & \makecell{89.1  ({\tiny 895/1{,}004})}
        & \makecell{\textbf{11.4}  ({\tiny 277/2{,}434})}
        & \makecell{75.3  ({\tiny 1{,}287/1{,}709})}
        & \makecell{73.5  ({\tiny 1{,}387/1{,}886})}
        & \makecell{86.9  ({\tiny 806/927})} \\
      & Ruby
        & \makecell{90.2  ({\tiny 1{,}161/1{,}287})}
        & \makecell{90.7  ({\tiny 1{,}180/1{,}301})}
        & \makecell{\textbf{69.8}  ({\tiny 981/1{,}405})}
        & \makecell{93.0  ({\tiny 745/801})}
        & \makecell{82.5  ({\tiny 1{,}053/1{,}276})}
        & \makecell{91.5  ({\tiny 1{,}355/1{,}481})}
        & \makecell{90.3  ({\tiny 1{,}058/1{,}172})}
        & \makecell{88.5  ({\tiny 711/803})} \\
      & Rust
        & \makecell{49.7  ({\tiny 1{,}016/2{,}046})}
        & \makecell{49.5  ({\tiny 1{,}014/2{,}050})}
        & \makecell{\textbf{38.6}  ({\tiny 924/2{,}393})}
        & \makecell{87.1  ({\tiny 681/782})}
        & \makecell{55.9  ({\tiny 1{,}103/1{,}973})}
        & \makecell{47.9  ({\tiny 962/2{,}009})}
        & \makecell{51.3  ({\tiny 1{,}088/2{,}120})}
        & \makecell{88.5  ({\tiny 637/720})} \\
    \midrule

        \multirow{4}{*}{\texttt{llama-8b}}
      & JavaScript
        & \makecell{80.1  ({\tiny 1{,}328/1{,}657})}
        & \makecell{80.2  ({\tiny 1{,}235/1{,}539})}
        & \makecell{\textbf{43.4}  ({\tiny 836/1{,}925})}
        & \makecell{89.4  ({\tiny 680/761})}
        & \makecell{49.0  ({\tiny 95/194})}
        & \makecell{77.4  ({\tiny 1{,}125/1{,}454})}
        & \makecell{72.2  ({\tiny 1{,}324/1{,}833})}
        & \makecell{89.6  ({\tiny 600/670})} \\
      & Python
        & \makecell{71.4  ({\tiny 1{,}275/1{,}785})}
        & \makecell{71.2  ({\tiny 1{,}300/1{,}826})}
        & \makecell{\textbf{32.5}  ({\tiny 1{,}069/3{,}288})}
        & \makecell{89.2  ({\tiny 815/914})}
        & \makecell{60.8  ({\tiny 818/1{,}345})}
        & \makecell{72.2  ({\tiny 1{,}212/1{,}678})}
        & \makecell{66.4  ({\tiny 1{,}515/2{,}283})}
        & \makecell{86.6  ({\tiny 720/831})} \\
      & Ruby
        & \makecell{90.4  ({\tiny 1{,}746/1{,}931})}
        & \makecell{87.2  ({\tiny 1{,}487/1{,}705})}
        & \makecell{\textbf{62.4}  ({\tiny 1{,}163/1{,}864})}
        & \makecell{94.6  ({\tiny 955/1{,}009})}
        & \makecell{83.9  ({\tiny 1{,}108/1{,}321})}
        & \makecell{90.1  ({\tiny 1{,}755/1{,}947})}
        & \makecell{85.8  ({\tiny 1{,}855/2{,}163})}
        & \makecell{94.1  ({\tiny 604/642})} \\
      & Rust
        & \makecell{49.1  ({\tiny 1{,}205/2{,}456})}
        & \makecell{48.5  ({\tiny 1{,}189/2{,}452})}
        & \makecell{\textbf{32.2}  ({\tiny 1{,}537/4{,}771})}
        & \makecell{81.6  ({\tiny 703/861})}
        & \makecell{71.4  ({\tiny 769/1{,}077})}
        & \makecell{48.4  ({\tiny 1{,}120/2{,}313})}
        & \makecell{49.7  ({\tiny 1{,}422/2{,}859})}
        & \makecell{51.0  ({\tiny 616/1{,}209})} \\
    \midrule

    \multirow{4}{*}{\texttt{mistral-7b}}
      & JavaScript
        & \makecell{65.7  ({\tiny 992/1{,}509})}
        & \makecell{66.1  ({\tiny 979/1{,}482})}
        & \makecell{\textbf{34.2}  ({\tiny 465/1{,}359})}
        & \makecell{79.2  ({\tiny 449/567})}
        & \makecell{62.1  ({\tiny 958/1{,}542})}
        & \makecell{64.8  ({\tiny 991/1{,}529})}
        & \makecell{68.1  ({\tiny 1{,}173/1{,}723})}
        & \makecell{82.7  ({\tiny 335/405})} \\
      & Python
        & \makecell{65.6  ({\tiny 1{,}553/2{,}367})}
        & \makecell{66.3  ({\tiny 1{,}573/2{,}374})}
        & \makecell{48.7  ({\tiny 696/1{,}429})}
        & \makecell{90.2  ({\tiny 826/916})}
        & \makecell{\textbf{41.5}  ({\tiny 832/2{,}006})}
        & \makecell{64.7  ({\tiny 1{,}505/2{,}327})}
        & \makecell{65.6  ({\tiny 1{,}575/2{,}402})}
        & \makecell{87.3  ({\tiny 738/845})} \\
      & Ruby
        & \makecell{76.6  ({\tiny 1{,}220/1{,}593})}
        & \makecell{80.9  ({\tiny 1{,}157/1{,}430})}
        & \makecell{\textbf{55.0}  ({\tiny 834/1{,}517})}
        & \makecell{86.8  ({\tiny 191/220})}
        & \makecell{79.2  ({\tiny 1{,}127/1{,}423})}
        & \makecell{82.3  ({\tiny 827/1{,}005})}
        & \makecell{85.0  ({\tiny 1{,}447/1{,}702})}
        & \makecell{93.7  ({\tiny 164/175})} \\
      & Rust
        & \makecell{46.6  ({\tiny 1{,}361/2{,}920})}
        & \makecell{46.1  ({\tiny 1{,}363/2{,}957})}
        & \makecell{\textbf{42.0}  ({\tiny 939/2{,}238})}
        & \makecell{81.3  ({\tiny 655/806})}
        & \makecell{56.9  ({\tiny 2{,}157/3{,}792})}
        & \makecell{46.7  ({\tiny 1{,}364/2{,}918})}
        & \makecell{50.4  ({\tiny 1{,}303/2{,}586})}
        & \makecell{77.3  ({\tiny 559/723})} \\ \midrule
        
    \multirow{4}{*}{\texttt{qwen-1.5b}}
      & JavaScript
        & \makecell{65.4  ({\tiny 981/1{,}500})}
        & \makecell{63.1  ({\tiny 1{,}040/1{,}647})}
        & \makecell{55.6  ({\tiny 789/1{,}418})}
        & \makecell{79.6  ({\tiny 565/710})}
        & \makecell{\textbf{46.9}  ({\tiny 927/1{,}975})}
        & \makecell{72.5  ({\tiny 1{,}173/1{,}619})}
        & \makecell{75.9  ({\tiny 1{,}438/1{,}894})}
        & \makecell{75.9  ({\tiny 183/241})} \\
      & Python
        & \makecell{71.6  ({\tiny 1{,}437/2{,}007})}
        & \makecell{74.9  ({\tiny 1{,}398/1{,}866})}
        & \makecell{62.6  ({\tiny 696/1{,}112})}
        & \makecell{85.1  ({\tiny 730/858})}
        & \makecell{\textbf{24.5}  ({\tiny 624/2{,}551})}
        & \makecell{81.0  ({\tiny 1{,}662/2{,}051})}
        & \makecell{76.3  ({\tiny 1{,}527/2{,}000})}
        & \makecell{77.1  ({\tiny 592/768})} \\
      & Ruby
        & \makecell{81.4  ({\tiny 1{,}265/1{,}554})}
        & \makecell{84.4  ({\tiny 1{,}488/1{,}763})}
        & \makecell{83.5  ({\tiny 988/1{,}183})}
        & \makecell{89.3  ({\tiny 507/568})}
        & \makecell{\textbf{64.2}  ({\tiny 1{,}688/2{,}629})}
        & \makecell{86.1  ({\tiny 1{,}112/1{,}291})}
        & \makecell{92.2  ({\tiny 1{,}921/2{,}084})}
        & \makecell{81.7  ({\tiny 384/470})} \\
      & Rust
        & \makecell{51.4  ({\tiny 1{,}639/3{,}186})}
        & \makecell{51.1  ({\tiny 1{,}624/3{,}175})}
        & \makecell{\textbf{29.8}  ({\tiny 941/3{,}159})}
        & \makecell{49.8  ({\tiny 696/1{,}398})}
        & \makecell{53.7  ({\tiny 1{,}380/2{,}569})}
        & \makecell{45.6  ({\tiny 1{,}182/2{,}592})}
        & \makecell{48.9  ({\tiny 1{,}192/2{,}437})}
        & \makecell{36.5  ({\tiny 519/1{,}423})} \\
    \midrule

    \multirow{4}{*}{\texttt{qwen-3b}}
      & JavaScript
        & \makecell{75.0  ({\tiny 1{,}227/1{,}636})}
        & \makecell{74.9  ({\tiny 1{,}199/1{,}600})}
        & \makecell{54.2  ({\tiny 623/1{,}150})}
        & \makecell{90.6  ({\tiny 903/997})}
        & \makecell{\textbf{50.5}  ({\tiny 963/1{,}908})}
        & \makecell{72.9  ({\tiny 1{,}193/1{,}636})}
        & \makecell{74.0  ({\tiny 1{,}321/1{,}785})}
        & \makecell{86.9  ({\tiny 637/733})} \\
      & Python
        & \makecell{74.2  ({\tiny 1{,}309/1{,}765})}
        & \makecell{76.6  ({\tiny 1{,}197/1{,}562})}
        & \makecell{38.3  ({\tiny 362/944})}
        & \makecell{86.3  ({\tiny 816/945})}
        & \makecell{\textbf{34.5}  ({\tiny 688/1{,}994})}
        & \makecell{72.5  ({\tiny 1{,}182/1{,}630})}
        & \makecell{68.1  ({\tiny 1{,}519/2{,}232})}
        & \makecell{80.0  ({\tiny 427/534})} \\
      & Ruby
        & \makecell{89.2  ({\tiny 1{,}736/1{,}947})}
        & \makecell{91.8  ({\tiny 1{,}813/1{,}974})}
        & \makecell{\textbf{57.9}  ({\tiny 987/1{,}704})}
        & \makecell{94.7  ({\tiny 1{,}323/1{,}397})}
        & \makecell{65.9  ({\tiny 1{,}328/2{,}016})}
        & \makecell{89.8  ({\tiny 1{,}880/2{,}093})}
        & \makecell{87.3  ({\tiny 2{,}096/2{,}400})}
        & \makecell{94.0  ({\tiny 864/919})} \\
      & Rust
        & \makecell{51.6  ({\tiny 1{,}424/2{,}758})}
        & \makecell{52.1  ({\tiny 1{,}461/2{,}803})}
        & \makecell{\textbf{44.3}  ({\tiny 947/2{,}139})}
        & \makecell{71.3  ({\tiny 536/752})}
        & \makecell{61.0  ({\tiny 904/1{,}483})}
        & \makecell{49.5  ({\tiny 1{,}188/2{,}401})}
        & \makecell{53.6  ({\tiny 1{,}546/2{,}886})}
        & \makecell{64.7  ({\tiny 389/601})} \\
    \bottomrule
    \end{tabular}
  \end{table*}

\begin{tcolorbox}[colback=gray!5, colframe=gray!50, title=\textbf{RQ4 Summary}]
No defense is universal under adversarial prompting. Adversarial prompts substantially increase PHR.
Ruby is consistently the most vulnerable language, while Rust is the most resistant. RAG and Self-Refine achieve the lowest PHR in 29 of 32 configurations, with RAG favoring lower-capacity models and Self-Refine stronger models. Decoding-only defenses remain unreliable under adversarial prompting.
\end{tcolorbox}
\section{Limitations and Threats to Validity}
\label{sec:limitations}

% We discuss the limitations and threats to validity.

\noindent
\textbf{Benchmark realism and generalizability.}
We extended the evaluation dataset from \cite{spracklen_we_nodate} to Ruby and Rust synthetically, with coding instructions generated by GPT-4o-mini from registry (package\_name, description) pairs. However, no human validation or comparison against authentic developer queries such as those found on Stack Overflow, GitHub Issues, or coding-assistant logs has been conducted. Nonetheless, we believe this controlled construction provides consistent coverage across languages and enables reproducible comparisons.

\noindent
\textbf{Model and ecosystem coverage.}
Our evaluation covers eight small open-weight models from five families and four programming-language ecosystems. This focus provides a challenging and practically relevant setting, but the findings may not generalize to larger models, other programming languages, or private  registries.

\noindent
\textbf{Configuration sensitivity.}
We apply one fixed hyperparameter configuration per strategy across all models and languages to support controlled comparison and do not perform optimization. Consequently, the reported rankings should be interpreted as results under the evaluated configurations rather than as upper bounds on each strategy's performance.

\section{Conclusion}
\label{sec:conclusion}

This paper revisited package hallucination in LLM-generated code from four angles: measurement, mitigation, utility, and adversarial robustness. Our corrected evaluation framework, which accounts for standard-library modules that prior pipelines misclassified as hallucinated, reveals that Python hallucination rates have been overstated by up to 9.4 percentage points in previous work. We recommend per-language standard-library exclusion as a baseline requirement for any future evaluation of package hallucination.

On the mitigation side, our systematic evaluation of five guided decoding strategies across eight models and four programming languages demonstrates that inference-time intervention is a viable and practical defense. Contrastive Decoding reduced hallucination in 19 of 32 configurations, while Nudging provided the best trade-off between hallucination reduction and output completeness, making it a promising lightweight strategy under standard, non-adversarial prompts among the guided decoding strategies. These results show that package hallucination can be substantially reduced without model retraining and without external retrieval infrastructure. Our utility analysis further shows that hallucination reduction and recommendation usefulness are not equivalent. Among all evaluated strategies, Greedy decoding provides the strongest average trade-off, followed by Self-Refine and Nudging.

At the same time, our adversarial evaluation exposes clear limits. Prompts seeded with fabricated package names amplify hallucination rates significantly by up to 58.1 percentage points, and all five decoding-only strategies fail to provide consistent protection under these conditions. RAG and Self-Refine performed best under hostile prompts, indicating protection requires external grounding or iterative self-verification. Practitioners should match defenses to the threat model: guided decoding for standard (non-adversarial) usage and stronger interventions for adversarial settings.
Future work should evaluate more sophisticated attack vectors, such as indirect prompt injection or multi-turn manipulation, and combining guided decoding with lightweight registry verification at inference time.

\section{Acknowledgment}
 
This work was supported by the Luxembourg Ministry of Foreign and European Affairs through
their Digital4Development (D4D) portfolio under the LuxWAyS project and the European Research
Council (ERC) under the European Union’s Horizon 2020 research and innovation program
(Project NATURAL - Grant agreement N° 949014).

\section*{Data Availability}
\label{sec:data}
The data used in this research are derived from publicly available sources. To ensure transparency and reproducibility, the source code, datasets, and usage instructions are publicly accessible in our anonymized repository at \url{https://zenodo.org/records/21786704}.

% \newpage
\bibliographystyle{ACM-Reference-Format}
\bibliography{main}

\end{document}